\documentclass[11pt,a4paper]{article}
\usepackage{amsmath}
\usepackage{amsfonts}
\usepackage{xcolor}
\usepackage{subfigure}
\usepackage{appendix}
\usepackage{amssymb}
\usepackage{hhline}
\usepackage{fancyhdr}
\usepackage{caption}
\usepackage{braket}
\usepackage{graphicx}
\usepackage[autocite    = superscript,
  backend     = bibtex,
  giveninits=true,
  sorting   = none,
  sortcites=true,
  style       = numeric-comp]{biblatex}
\usepackage{geometry}
\usepackage{color}
\usepackage{authblk}
\usepackage{float}
\usepackage{hyperref}

\DeclareNameAlias{author}{last-first}

\title{\textbf{\huge Cosmology at the end of the world}}
\author[1,*]{\Large Stefano Antonini}
\author[1,2]{\Large Brian Swingle}
\affil[1]{\normalsize \textit{Maryland Center for Fundamental Physics and Department of Physics, University of Maryland, College Park MD 20742, USA}}
\affil[2]{\normalsize \textit{Condensed Matter Theory Center and Joint Center for Quantum Information and Computer Science, University of Maryland, College Park MD 20742, USA}}
\affil[*]{\textit{Email: santonin@umd.edu}}
\date{v1: July 17, 2019\\ v2 (this version): January 21, 2021}

\begin{document}

\maketitle

\begin{abstract}
In the last two decades the Anti-de Sitter/Conformal Field Theory correspondence (AdS/CFT) has emerged as focal point of many research interests. In particular, it functions as a stepping stone to a still missing full quantum theory of gravity. In this context, a pivotal question is if and how cosmological physics can be studied using AdS/CFT. Motivated by string theory, braneworld cosmologies propose that our universe is a four-dimensional membrane embedded in a bulk five-dimensional AdS spacetime. We show how such a scenario can be microscopically realized in AdS/CFT using special field theory states dual to an ``end-of-the-world brane'' moving in a charged black hole spacetime. Observers on the brane experience cosmological physics and approximately four-dimensional gravity, at least locally in spacetime. This result opens a new path towards a description of quantum cosmology and the simulation of cosmology on quantum machines.
\end{abstract}

\headsep = 30pt
\pagestyle{fancy}
\fancyhead[LE,RO]{}
\fancyhead[RE,LO]{\textbf{Cosmology at the end of the world}}
\fancyfoot[C]{\textbf \thepage}
\renewcommand{\headrulewidth}{0.4pt}
\renewcommand{\footrulewidth}{0pt}

\tableofcontents

\section{Introduction}

The achievement of a quantum mechanical description of spacetime is one of the most challenging issues facing modern physics. The Anti-de Sitter/Conformal Field Theory (AdS/CFT) correspondence provides a promising starting point to accomplish this goal. As a concrete realization of the holographic principle\cite{thooft,susskind}, it relates the observables of a full quantum gravity theory living in a ($d+1$)-dimensional spacetime (the bulk) to those of a $d$-dimensional dual field theory without gravity associated to the boundary of the bulk spacetime\cite{review}. However, at this stage AdS/CFT must still be regarded as a toy model of the physical world, since the asymptotically AdS spacetimes it describes are different from the universe we observe. In this paper we exhibit a viable way to study certain Friedmann-Lema\^itre-Robertson-Walker (FLRW) cosmologies using the AdS/CFT correspondence, realizing a recently proposed new perspective on braneworld cosmology and quantum gravity in a cosmological universe\cite{bhmicrostate}.

\begin{figure}[H]
\centering
\includegraphics[scale=0.3]{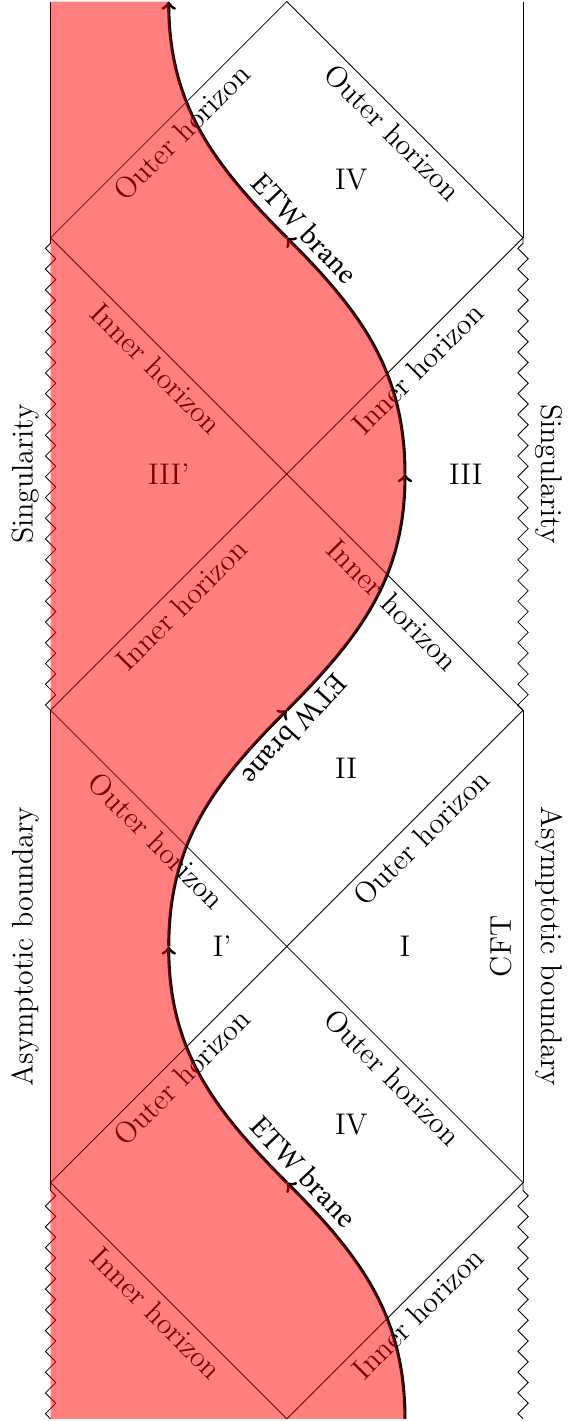}
\caption{\textbf{Brane trajectory.} Maximally extended Penrose diagram for the AdS-Reissner-Nordstr\"om black hole. Only the time and radial directions are represented, while the other spatial dimensions are suppressed. Light rays move at an angle of $45^\circ$. More patches of the AdS-RN spacetime are glued together here. Region I is the exterior of the black hole and the CFT lives on the asymptotic boundary associated with it. Region I' is the corresponding second asymptotic region, typical of maximally extended black holes. In Regions II and IV, which are bounded by the outer and inner horizons, the radial coordinate is timelike, while the time coordinate is spacelike. Regions III and III' are the interior regions of the maximally extended black hole. The ETW brane oscillates inside and outside the two horizons of the black hole, cutting off the left asymptotic region I'. The red region, generally part of the maximally extended charged black hole, is not present in our spacetime.}
\label{penrose}
\end{figure}

AdS/CFT associates different states of the CFT to different geometries of the bulk spacetime. In particular, certain highly excited pure states of the CFT have a dual geometry hosting a black hole and a dynamical end-of-the-world (ETW) brane\cite{bound1,bound2,bound3,bound4,bound5}. This bulk configuration is built from a two-sided wormhole spacetime by replacing the left boundary with the ETW brane (see Figure \ref{penrose}). The original wormhole spacetime describes an entangled state of two CFTs, and the ETW brane corresponds to a complete measurement of one CFT to leave a pure state of the remaining CFT. Hence, observables in the CFT must probe the physics behind the horizon of the black hole, including the remaining part of the left asymptotic region\cite{bth1,bth2,bth3,bth4} and the evolution of the ETW brane\cite{hartmalda,bhmicrostate}.

Although we do not know how to do the decoding in general, some CFT observables have been explicitly shown to probe the physics behind the horizon in similar setups. The entanglement entropy of large regions of the CFT\cite{hartmalda,bhmicrostate} and the holographic complexity\cite{bhmicrostate} exhibit a time-dependent behaviour containing information about the evolution of the ETW brane and its tension. Although we will not evaluate explicitly the mentioned quantities in this paper, they can be calculated in our setup by directly generalizing the techniques employed in ref\cite{bhmicrostate}.
\noindent The presence of the brane opens an interesting scenario: when the bulk spacetime is 5-dimensional, an observer living on the 4-dimensional ETW brane would interpret the motion of the brane as the evolution of a FLRW universe, where the radial position plays the role of the scale factor\cite{friedmann,padilla1,padilla2,cosmoholors1,cosmoholors2,cosmoholors3}. Hence, holographic duality allows to describe such ``braneworld cosmologies'' using CFT observables associated with the right asymptotic region of the black hole\cite{bhmicrostate}. Holographic cosmologies have been considered previously\cite{cosmoholo1,cosmoholo2,cosmoholo3}, also in the context of de Sitter holography\cite{dsds} and braneworld holography\cite{cosmoholors1,cosmoholors2,cosmoholors3,cosmoholors4}. However, in the latter cases the holographic CFT lives on the brane and is coupled to gravity, whereas we are proposing a non-perturbative CFT description of the entire spacetime including the brane.

Now, a braneworld cosmological model can only be realistic if, for an observer living on the brane, gravity is effectively 4-dimensional and localized on the brane. The Randall-Sundrum model\cite{RS1,RS2} provides a mechanism for gravity localization on 4-dimensional branes embedded in AdS$_5$ bulk, and the mechanism can be generalized to different brane geometries\cite{karchrandall} and to the presence of black holes\cite{resonances2,resonances}. If the brane is not too close to the black hole horizon, gravity is localized, but only locally in spacetime: an experimentalist on the brane would observe ordinary gravity only on non-cosmological scales, and for a limited range of time. In this work we show that, in the presence of a charged black hole in the bulk (AdS-Reissner-Nordstr\"om)\cite{dias,huang,2horizons,btz,RN,tempestr}, such gravity localization is achievable without losing the dual CFT description of the spacetime. This proof-of-principle result, not attainable in the simplest setup of an AdS-Schwarzschild black hole and a pure tension brane\cite{bhmicrostate}, opens the door to a new formulation of cosmology within AdS/CFT. Our choice of units is $\hbar=c=\varepsilon_0=k_B=1$.

\section{Euclidean analysis}

\label{euclsec}


The spacetime described above is dual to a CFT state that can be prepared in principle using a quantum computer. Theoretically, it is convenient to describe it starting from a Euclidean path integral. Extending previous works\cite{tfdrn,bhmicrostate}, the dual state is a fixed charge boundary state of the CFT on a spatial sphere $\mathbb{S}^{d-1}$ evolved for an amount $\tau_0$ of imaginary (Euclidean) time, called preparation time:
\begin{equation}
\ket{\Psi}=\textrm{e}^{-\tau_0 H}\ket{B,Q}.
\label{state}
\end{equation}
The ETW brane is associated to the past boundary condition in the path integral. The preparation time $\tau_0$ of the CFT state must be positive, because otherwise the state is not normalizable and hence not in the CFT Hilbert space.

\begin{figure}[h]
\centering
\includegraphics[height=4cm]{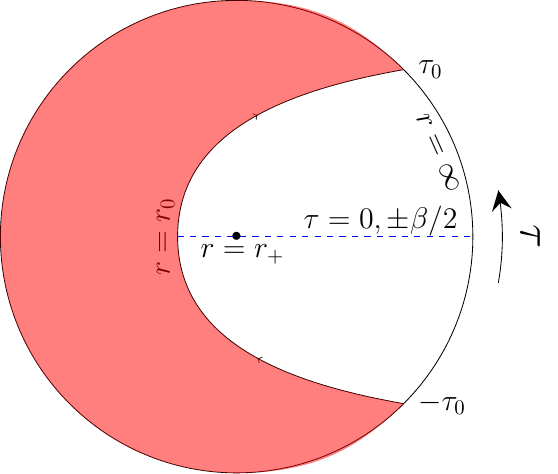}
\caption{\textbf{Euclidean brane trajectory.} The Euclidean path integral on this spacetime computes the norm of the CFT state. The radial coordinate is the radius $r$, with the outer horizon $r=r_+$ at the center and the asymptotic boundary $r=\infty$ at the circumference. The other spatial dimensions are suppressed. The angular coordinate is the Euclidean time $\tau$, which has a periodicity $\beta$. The blue dashed line represents the $\tau=0,\pm\beta/2$ slice, where the brane reaches its minimum radius $r_0$. The red region is cut off by the ETW brane. The portion of asymptotic boundary not excised by the brane gives twice the preparation time $\tau_0$ of the CFT state.}
\label{euclidean}
\end{figure}

In Euclidean signature, the brane is the bulk extension of the CFT boundary\cite{bound2,bound3,bhmicrostate}. Its trajectory starts at the asymptotic boundary $r=\infty$ at $\tau=-\tau_0$, reaches a minimum radius $r=r_0$ and ends again on the boundary at $\tau=\tau_0$. Because the state is highly excited, the bulk geometry will also typically contain a black hole. The total Euclidean periodicity is given by the inverse temperature $\beta$ of the black hole, determined by its size and charge. The brane trajectory excises part of this total range of Euclidean time, leaving only the interval $[-\tau_0,\tau_0]$ where the CFT is defined. Thus, the preparation time, which is a property of the CFT state we are considering, can be written in terms of bulk quantities as $2\tau_0=\beta-2\Delta\tau$ for positive-tension branes, where $\Delta\tau$ is the total Euclidean time needed for the brane to cover half of its trajectory. As noted above, the preparation time must be positive to have a sensible CFT state. Therefore, only the Euclidean geometries with a positive preparation time will be dual to a CFT state of the type described in equation (\ref{state})\cite{bhmicrostate}. We call such geometries ``sensible Euclidean solutions''. Alternatively, if $\tau\leq 0$, the brane overlaps itself and there is no Euclidean time interval $[-\tau_0,\tau_0]$ on the boundary where to define the CFT\cite{bhmicrostate} (see Figure \ref{euclidean}).

The bulk physics is specified by the Euclidean action, which for the Einstein-Maxwell system considered here is $I=I_{bulk}+I_{ETW}$. $I_{bulk}$ is the Einstein-Maxwell action with a negative cosmological constant $\Lambda=-d(d-1)/(2L^2_{AdS})$, including a Gibbons-Hawking-York term and an electromagnetic boundary term (needed if the charge is held fixed\cite{hawkingboundary,1-,phase}) for the asymptotic boundary. The brane action reads:
\begin{equation}
I_{ETW}=-\frac{1}{8\pi G}\int_{ETW}d^dx\sqrt{h}\left[K-(d-1)T\right]+I^{em}_{ETW},
\label{braneaction}
\end{equation}
where $G$ is the $(d+1)$-dimensional Newton constant, $K$ is the trace of the extrinsic curvature of the brane, $T$ is the tension, $h$ is the determinant of the metric induced on the brane and $I^{em}_{ETW}$ is an electromagnetic boundary term. The variation of the bulk action leads to the Einstein-Maxwell equations with a negative cosmological constant, whose solution is the AdS-Reissner-Nordstr\"om (AdS-RN) metric (in Euclidean signature):
\begin{equation}
ds^2=f(r)d\tau^2+\frac{dr^2}{f(r)}+r^2d\Omega_{d-1}^2
\label{linel}
\end{equation}
with (for $d>2$)
\begin{equation}
f(r)=1+\frac{r^2}{L^2_{AdS}}-\frac{2\mu}{r^{d-2}}+\frac{Q^2}{r^{2(d-2)}}
\label{four}
\end{equation}
where $\mu$ and $Q$ are the mass and charge parameters of the black hole. They can be expressed as a function of the inner (Cauchy) and outer (event) horizons of the black hole, solutions of the equations $f(r_+)=f(r_-)=0$. The black hole is called extremal when $r_-=r_+$, corresponding to the maximum admissible charge for a fixed mass. 

Varying the brane action and imposing Neumann boundary conditions\cite{bound2,bound3}, the equation
\begin{equation}
K_{ab}-Kh_{ab}=(1-d)Th_{ab}
\label{brane}
\end{equation}
is obtained, with $a,b=0,...,d-1$. By parametrizing the trajectory of the brane with $r=r(\tau)$, the $\tau\tau$ component of equation (\ref{brane}) yields
\begin{equation}
\frac{dr}{d\tau}=\pm \frac{f(r)}{Tr}\sqrt{f(r)-T^2r^2}
\label{traj}
\end{equation}
where the sign depends on if the brane is expanding or contracting. The other components of equation (\ref{brane}) give an equation for $d^2r/d\tau^2$, which can also be obtained from equation (\ref{traj}) (more details are reported in Appendix \ref{appendixb}). For $d>2$ and $|T|<T_{crit}=1/L_{AdS}$, the equation $f(r)=T^2r^2$, which gives the zeros of equation (\ref{traj}), has always two solutions (which can be computed numerically): $r_0^+>r_+$ and $r_0^-<r_-$. Between the two solutions the square root takes imaginary values. Since the brane is contracting from and expanding to $r=\infty$, the minimum radius that it reaches at the inversion point is $r_0=r_0^+$ and is always outside the event horizon $r_+$. The role of the solution $r_0^-$ will be clear later. Since we are interested in time-symmetric solutions, we require $r_0$ to sit on the $\tau=0,\pm\beta/2$ line. In this paper we will consider only positive tensions, corresponding to the geometry in Figure \ref{euclidean}.

For negative tension branes, the part of geometry retained is the one shaded in red in Figure \ref{euclidean} and does not include the black hole horizon. Furthermore, the preparation time in such a case is defined as $\tau_0=\Delta\tau$\cite{bound3}. Although these solutions are interesting, it is not clear if they correspond to a dual CFT state\cite{bhmicrostate} and if they can support gravity localization. Therefore we leave their study to future work. From the CFT point of view, this is justified because we can set the charge, preparation time, and boundary condition via the path integral. The boundary condition determines the tension, hence we may restrict to positive tension.
\begin{figure}[h]
\centering
\includegraphics[width=\textwidth]{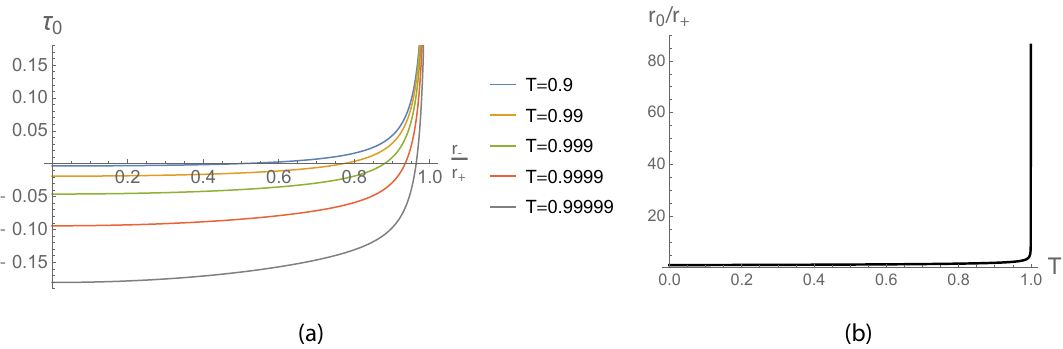}
\caption{\textbf{Euclidean sensible solutions.} (a) Preparation time $\tau_0$ as a function of the ratio between inner and outer horizon radii $r_-/r_+$ for different values of the tension $T$. The larger the ratio, the closer the black hole to extremality. The preparation time is positive for every value of the tension if the black hole is sufficiently close to extremality. We chose four spatial dimensions ($d=4$), $r_+=100$ and we set the AdS radius to be $L_{AdS}=1$. (b) The ratio between the minimum radius of the brane $r_0$ and the outer horizon radius $r_+$ grows with the tension $T$. Our choice of parameters is $d=4$, $r_+=100$, $r_-=99.9$, $L_{AdS}=1$. When the brane is near-critical ($T\to 1/L_{AdS}$), it is far from the black hole horizon at the inversion point, and in the Lorentzian picture gravity can be locally localized on it.}
\label{fig3}
\end{figure}
Given values of $d$, $r_+$, $r_-$ and $T$, it is possible to evaluate the Euclidean time $\Delta\tau$ and therefore the preparation time $\tau_0$ (see Appendix \ref{appendixb}). We find that, for a black hole sufficiently close to extremality, i.e. for $r_-\to r_+$, it is possible to approach the critical value of the tension $T_{crit}$ while retaining a sensible Euclidean solution, i.e. with $\tau_0>0$. At the same time, when $T\to T_{crit}$ the ratio between the minimum radius of the brane and the outer horizon radius becomes very large. We further remark (see Appendix \ref{actcomp}) that this non-extremal black hole solution is always dominant in the thermodynamic ensemble when $\tau_0>0$. This result was not feasible in the AdS-Schwarzschild case\cite{bhmicrostate} and is first main result of this work.

As noted above, the CFT states we consider can be prepared given full quantum control of the CFT. Euclidean evolution is not unitary and cannot be deterministically implemented. However, by preparing the thermofield double state\cite{eternal_wormhole,building_tfd,product_spectrum} and measuring the left side in a simple basis\cite{bound4}, a random such state can be prepared.

\section{Lorentzian analysis and braneworld cosmology}

In order to study cosmological physics, we must find the Lorentzian geometry associated with the Euclidean picture outlined above. Consider the state (\ref{state}), obtained taking the $\tau=0,\pm\beta/2$ slice of the thermal circle (where the brane reaches its minimum radius $r_0$), and evolve it in real time. The corresponding geometry is a maximally extended AdS-RN black hole with the ETW brane cutting off part of the left asymptotic region. The effect of the transition to Lorentzian signature on equation (\ref{traj}) is to flip the sign of the radicand, thus the RHS is now real only if $r_0^-<r<r_0^+$. The minimum radius in Euclidean signature is a maximum radius in Lorentzian signature. Therefore the brane expands and contracts crossing the two horizons of the black hole, but, differently from the AdS-Schwarzschild case, it never reaches the singularity (see Figure \ref{penrose}).

Defining the brane proper time $d\lambda^2=[f(r)-r'^2/f(r)]dt^2$, the metric induced on the brane takes the form
\begin{equation}
ds_{ETW}^2=-d\lambda^2+r^2(\lambda)d\Omega_{d-1}^2
\end{equation}
which is a closed FLRW metric where the brane radius $r(\lambda)$ plays the role of the scale factor, ranging between $r_0^-$ and $r_0^+$ and satisfying the Friedmann equation
\begin{equation}
\left(\frac{\dot{r}}{r}\right)^2=-\frac{1}{r^2}+\frac{2\mu}{r^d}-\frac{Q^2}{r^{2d-2}}+\left(T^2-\frac{1}{L^2_{AdS}}\right)
\label{friedmain}
\end{equation}
where the dot indicates a derivative with respect to the proper time. For $d=4$, this result (already obtained in refs.\cite{padilla2,cosmoholors3}) shows that an observer comoving with the brane interprets the motion of the brane in the bulk as the expansion and contraction of a closed FLRW universe in the presence of radiation (with energy density proportional to the mass of the black hole), a negative cosmological constant $\Lambda_4=3(T^2-1/L^2_{AdS})$ (which is very small when $T\to T_{crit}$) and stiff matter\cite{stiff,cosmoholors3} with negative energy density (proportional to the charge of the black hole). 

From the bulk point of view, the existence of a minimum radius is due to the repulsive nature of the RN singularity, while from the braneworld point of view, the (repulsive) stiff matter term is responsible for the absence of the cosmological singularity, replaced by a Big Bounce\cite{stiff}. In principle, gluing together more patches of the AdS-RN bulk spacetime, we can obtain a cyclical cosmology. However, the instability of the Cauchy horizon against even small perturbations suggests that, in the region near the bounce, the present description of the brane evolution is not reliable.\cite{instability} When the charge of the black hole vanishes, we recover the AdS-Schwarzschild description of a Big Bang-Big Crunch braneworld cosmology. We emphasize that the cosmological interpretation is not appropriate in the region where the Big Bounce takes place, because a 4-dimensional description of gravity localized on the brane is achievable only when the brane sits far from the black hole horizon.

\section{Gravity localization}
\label{gravloc}

The remaining issue is whether observers on the brane see approximately four-dimensional gravity. Randall and Sundrum showed\cite{RS1,RS2} that a 4-dimensional Minkowski brane embedded in a warped AdS$_5$ spacetime supports a normalizable graviton zero-mode bound on the brane, reproducing 4-dimensional gravity. The bound mode is lost in the presence of an AdS-Schwarzschild black hole in the bulk\cite{resonances2}, but a resonant quasi-bound mode with a finite lifetime persists when the brane is static and far from the black hole horizon\cite{resonances}. This is a metastable mode that, after a finite lifetime, ``leaks'' into the bulk and falls into the black hole horizon. The shorter the spatial scale of the gravitational perturbation, the longer it is bound on the brane. Gravity is therefore locally localized both in space and in time, meaning that it will look 4-dimensional to an observer living on the brane, but only on spatial and time scales smaller than the cosmological ones.

Clarkson and Seahra\cite{resonances} focused their attention mainly on the small black hole case ($r_H<L_{AdS}$). Since sensible Euclidean solutions with $r_0/r_+\gg 1$ are achievable only if the black hole is large, our generalization of their work to the AdS-RN spacetime is focused on the $r_+\gg L_{AdS}$ case (see Appendix \ref{gravlocdet}). 

Let us consider a linear perturbation of the metric $\delta g_{\mu\nu}=g_{\mu\nu}-g_{\mu\nu}^0$, $\mu,\nu=0,...,d$. As a tensor on a spatial slice at constant radius ($t=const$, $r=const$), it can be decomposed into scalar, vector and tensor components\cite{kodama}. The graviton mode of interest is the tensor component which has $\delta g_{t\mu}=\delta g_{r\mu}=0$. We also use the transverse-traceless (TT) gauge condition $\delta g^\mu_\mu=0=\nabla^\mu\delta g_{\mu\nu}$. It is useful to introduce the adimensional coordinates and parameters $y=r/r_+$, $\tilde{t}=t/r_+$, $\gamma=L_{AdS}/r_+$, $q=Q/r_+^{d-2}$ and to decompose the metric perturbation in terms of tensor harmonics $\mathbb{T}^{(k)}_{ij}$ (with $i,j=1,...,d-1$) on the unit ($d-1$)-sphere. They satisfy $\Delta_{d-1} \mathbb{T}^{(k)}_{ij}=-k^2\mathbb{T}^{(k)}_{ij}$, where $\Delta_{d-1}$ is the ($d-1$)-dimensional covariant Laplacian on the unit ($d-1$)-sphere and $k^2=l(l+d-2)-2$, with $l=1,2,...$ generalized angular momentum. 

Defining the tortoise coordinate $dr^*=dy/f(y)$ and studying the problem in the frequency domain, the linearized Einstein equations can be recast in the form of a one-dimensional Schr\"odinger equation for each tensor mode:
\begin{equation}
-\partial_{r^*}^2\psi_{k,\omega}(r^*)+V_k\left[y(r^*)\right]\psi_{k,\omega}(r^*)=\omega^2\psi_{k,\omega}(r^*)
\label{schro}
\end{equation}
where the potential reads:
\begin{equation}
V_k(y)=f(y)\left[\frac{d^2-1}{4\gamma^2}+\frac{d^2-4d+11+k^2}{4y^2}+\frac{(d-1)^2\left(1+\frac{1}{\gamma^2}+q^2\right)}{4y^d}-\frac{(d-1)(3d-5)q^2}{4y^{2d-2}}\right].
\label{potential}
\end{equation}
This potential diverges for $r^*=r^*_\infty$ (where $y(r^*_\infty)=\infty$) and vanishes exponentially for $r^*\to -\infty$, at the black hole horizon.
\begin{figure}[H]
\centering
\includegraphics[height=4cm]{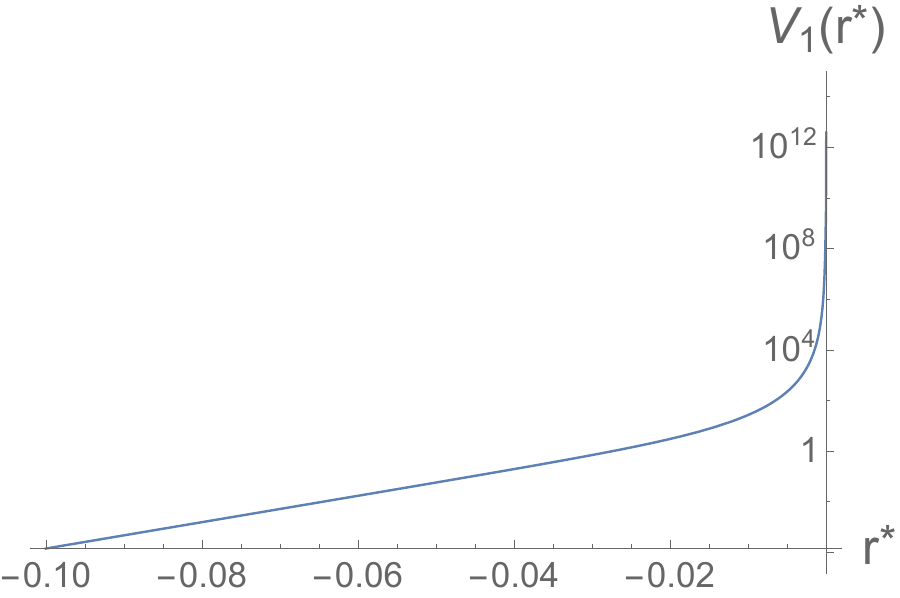}
\caption{\textbf{Potential $\mathbf{V_k[y(r^*)}$] - Large black hole.} Potential (\ref{potential}) as a function of the tortoise coordinate $r^*$ in four spatial dimensions ($d=4$). We chose the values $r_+=100$, $r_-=99.9$ for the outer and inner horizon radii respectively, setting the AdS length to be $L_{AdS}=1$ and the angular momentum (and therefore the eigenvalue $k$) to be $l=k=1$. The potential diverges for $r^*_\infty=4.94\cdot 10^{-5}$ and vanishes exponentially at the horizon $r^*\to -\infty$.}
\label{fig4}
\end{figure}
\noindent Now we need to find a boundary condition on the ETW brane, which cuts off the radial coordinate at $r^*=r^*_b<r^*_\infty$. If the brane is moving in the bulk, this is a non-trivial task. We assume that the brane is moving adiabatically with respect to the time scale of the perturbation, and consider it effectively static at a position $r^*=r^*_b$, verifying later the reliability of the adiabatic approximation. Under this assumption, the linearized version of equation (\ref{brane}) provides the (Neumann) boundary condition $\partial_{r^*}\psi_{k,\omega}|_{r^*=r^*_b}=(d-1)f(y)/(2y)\cdot \psi_{k,\omega}|_{r^*=r^*_b}$, with $y=y(r^*)$. Requiring this boundary condition is equivalent to add to the potential (\ref{potential}) a negative delta at the position of the brane, whose depth is $(d-1)f(y)/y$. In the original Randall-Sundrum model, such a delta guarantees the existence of the zero bound mode. The rapid vanishing behaviour of the potential (\ref{potential}) at the black hole horizon implies that only a metastable, quasi-bound mode with complex frequency $\omega=\bar{\omega}+i\Gamma/2$ and pure infalling boundary condition at the black hole horizon (typical of quasi-normal modes) can be present in our setup. In the large black hole case ($\gamma\ll 1$) this is the only resonant mode\cite{resonances}.

\begin{figure}[h]
\centering
\includegraphics[width=\textwidth]{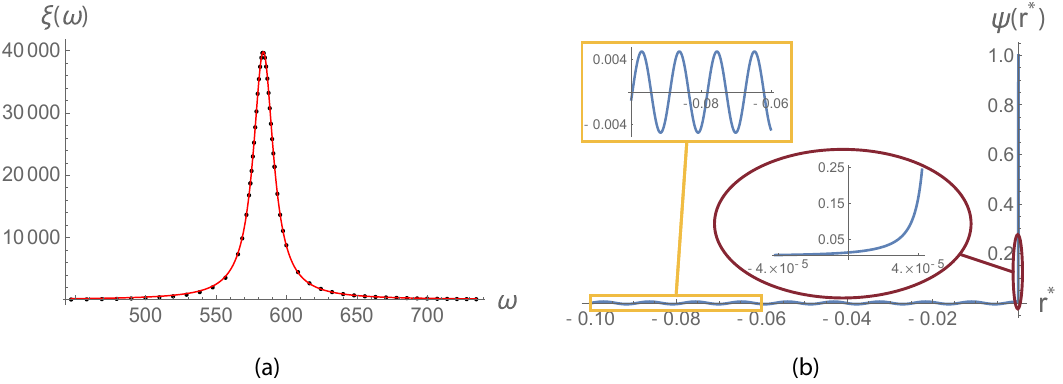}
\caption{\textbf{Quasi-bound mode.} Results for four spatial dimensions ($d=4$), with outer and inner horizon radii given by $r_+=100$ and $r_-=99.9$ respectively. We set the AdS length to be $L_{AdS}=1$, giving $\gamma=L_{AdS}/r_+=0.01\ll 1$. The choice for the angular momentum is $l=5$, while the position of the brane is taken to be $y_b=34.62$ (corresponding to the maximum radius of a brane with tension $T=0.999999$). (a) Squared trapping coefficient $\xi(\omega)$ as a function of frequency $\omega$. $\xi(\omega)$ presents a Breit-Wigner peak centered at the real part $\bar{\omega}$ of the frequency of the quasi-bound mode, while its imaginary part $\Gamma/2$ is given by the half-width at half-maximum. The fit gives $\bar{\omega}=583.6$, $\Gamma=17.28$. The adjusted coefficient of determination for the fit is $R^2=0.999978$, showing the accuracy of the Breit-Wigner approximation for the peak of the squared trapping coefficient for this choice of parameters. (b) Quasi-bound mode wavefunction $\psi(r^*)$ as a function of the tortoise coordinate $r^*$. $\psi(r^*)$ oscillates in the near-horizon region, as shown in the top-left inset, and grows very fast near $r^*_b=4.65\cdot 10^{-5}$, as it is evident from the right-bottom inset. Therefore, the quasi-bound mode is very well localized on the brane, but has a non-vanishing probability to leak into the bulk and fall into the black hole.}
\label{mode}
\end{figure}
Focusing on the $d=4$, $\gamma\ll 1$ case of interest, we use the trapping coefficient method\cite{resonances} (see Appendix \ref{gravlocdet}) to find the real and imaginary parts of the frequency of the quasi-bound mode. For a given value of the charge of the black hole, there are three control parameters: the size of the black hole $\gamma$, the position of the brane $r^*(y_b)$, and the angular momentum $l$. We find that, when the brane is sufficiently far from the horizon of the black hole, $\bar{\omega}\gg\Gamma$ (i.e. the mode is long-lived) and the graviton is very well localized on the brane (see Figure \ref{mode}). Such a localization is lost if the brane radius becomes too large. Increasing the size of the black hole, the brane must sit farther from the horizon to retain the long-lived quasi-bound mode. Gravity localization is also more efficient for higher values of the angular momentum $l$. Increasing $l$ means probing shorter distances on the brane. Therefore, this result is in accordance with our expectation to obtain a 4-dimensional effective description of gravity only locally.
\begin{table}[h]
\centering
\scalebox{0.9}{
\begin{tabular}{|c|c|c|c|c|c|}
$y_b$ &$T_H$ &$t_o$ &$t_d$ &$\left(y'\right)^2/\left[f(y)\right]^2$ &$\lambda(y_b)/\lambda_{tot}$\\ \hline
30 &$1.762\cdot 10^{-3}$ &$9.172\cdot 10^{-4}$ &0.08370 &$3.581\cdot 10^{-6}$ &0.9764 \\ \hline
45 &$2.698\cdot 10^{-3}$ &$9.190\cdot 10^{-4}$ &0.1880 &$6.786\cdot 10^{-7}$ &0.9465 \\ \hline
60 &$3.716\cdot 10^{-3}$ &$9.229\cdot 10^{-4}$ &0.3342 &$2.012\cdot 10^{-7}$ &0.9035 \\ \hline
75 &$4.875\cdot 10^{-3}$ &$9.282\cdot 10^{-4}$ &0.5226 &$7.484\cdot 10^{-8}$ &0.8459 \\ \hline
90 &$6.282\cdot 10^{-3}$ &$9.350\cdot 10^{-4}$ &0.7516 &$3.129\cdot 10^{-8}$ &0.7713 \\ \hline
105 &$8.178\cdot 10^{-3}$ &$9.432\cdot 10^{-4}$ &1.026 &$1.356\cdot 10^{-8}$ &0.6748 \\ \hline
120 &$0.01124$ &$9.530\cdot 10^{-4}$ &1.339 &$5.495\cdot 10^{-9}$ &0.5462 \\ \hline
135 &$0.01895$ &$9.645\cdot 10^{-4}$ &1.713 &$1.527\cdot 10^{-9}$ &0.3540 \\ \hline
145.1 (max) &$\infty$ &$9.733\cdot 10^{-4}$ &1.956 &0 &0 \\ \hline
\end{tabular}}
\caption{\textbf{Adiabatic approximation: time scales comparison.} Results for four spatial dimensions ($d=4$) and a large, near-extremal black hole with outer and inner horizon radii given by $r_+=100$ and $r_-=99.9$ respectively. We set the AdS radius to be $L_{AdS}=1$ (i.e. $\gamma=L_{AdS}/r_+=0.01$). We further chose the value $l=10$ for the angular momentum and $T=0.999999999$ for the brane tension, which is clearly near-critical. With this choice of parameters, the maximum radius reached by the brane is $y_b^{max}=145.1$ (reported in the last line of the first column). In the first column we show the position of the brane in adimensional coordinates $y_b=r_b/r_+$. In the second, third and fourth columns the time scales of motion of the brane ($T_H$) and of oscillation ($t_o$) and decay ($t_d$) of the quasi-bound mode are reported, respectively. When $T_H\gg t_o$ is satisfied (for example in the last three lines), the adiabatic approximation holds. If $t_d\gg T_H$, the adiabatic approximation breaks down before the mode leaks into the bulk. The fifth column shows the ratio $\left(y'\right)^2/\left[f(y)\right]^2$, where $y'=dy/dt$ and $f(y)$ is the absolute value of the 00 component of the metric (given by equation (\ref{four})) expressed in terms of the adimensional coordinates and parameters. This ratio must be small in order for our analysis to be consistent. In the last column we report the fraction between the proper time $\lambda(y_b)$ needed for the brane to expand from $y_b$ to $y_b^{max}$ and shrink back to $y_b$, and the total amount of proper time needed to complete the brane trajectory. The Euclidean solution corresponding to this choice of parameters is sensible and dominant in the thermodynamic ensemble.}
\label{timescales}
\end{table}
The last step is to verify that the adiabatic approximation is reasonable. One must compare the time scale of oscillation $t_o=1/\bar{\omega}$ with the ``Hubble time'' $T_H=y(t)/y'(t)$, determined by the Lorentzian version of equation (\ref{traj}). The adiabatic approximation is reliable if $t_o\ll T_H$ and if the condition $(y')^2/f^2(y)\ll 1$ (used to derive the boundary condition at the brane, see Appendix \ref{appendixe1}) is satisfied. For given values of charge and tension, the approximation is arbitrarily good when approaching the inversion point in the trajectory of the brane, where $y'$ vanishes. But it holds also for a significant portion of the trajectory of the brane, as quantified by the ratio between the proper time spent to cover the part of trajectory where the approximation is reliable and the total amount of proper time for the entire trajectory. As an example, for the set of parameters considered in Table \ref{timescales}, if we accept $t_o\sim T_H/10$ as the threshold for the validity of the adiabatic assumption, the latter holds for $\sim 55\%$ of the total amount of brane proper time. The decay time $t_d=2/\Gamma$ is almost always considerably larger than the Hubble time, meaning that our analysis loses significance before the quasi-bound mode leaks into the bulk. We finally remark that the time of oscillation of the mode is very close to the one expected for a 4-dimensional metric perturbation of the Einstein-static universe $t_{GR}=y_b/\sqrt{f(y_b)(l+2)l}$. The local localization of gravity on the Euclidean-sensible braneworld solution is the second main result of this work.

\section{Discussion}

In this work we proved that it is possible to build a holographic braneworld cosmology model which admits locally a 4-dimensional effective description of gravity. An interesting open question is if this framework can be extended to different cosmological models, eventually including a de Sitter phase that could match our current observational data, and if gravity localization is achievable on even larger scales in such setups. Understanding gravity localization beyond the adiabatic approximation would also be interesting.

The holographic dual description presents many non-trivial issues to be explored. Determining which CFT observables are able to probe the physics behind the horizon of the black hole, and therefore the braneworld cosmology, is crucial. If the brane is not too far from the black hole horizon, one possibility is the entanglement entropy of large spatial regions of the CFT\cite{bhmicrostate,rt1,rt2}, while additional information can be extracted from the holographic complexity, even if its CFT interpretation is not completely clear yet. It is also important to identify field theories with the right boundary states to make our construction work and verify the stability of the solutions in the full theory. For example, one should check that scalar fields do not condense in the near-extremal black hole background (or that the desired properties persist in such a condensed state). Explicit realizations of AdS/BCFT in top-down stringy models have been proposed in refs\cite{bound3,topdown}. Given a CFT, it should be possible to simulate it on an quantum computer\cite{eternal_wormhole,building_tfd,product_spectrum,bound4}, thereby opening up the possibility to experimentally study holographic braneworld cosmology.

Although many questions still need to be answered, the model presented in this paper shows the possibility to holographically describe a FLRW cosmology using AdS/CFT, bringing us one step closer to the understanding of quantum gravity in a cosmological universe.

\section*{\small{Acknowledgements}} 

\addcontentsline{toc}{section}{Aknowledgements}

This work was supported in part by the U.S. Department of Energy, Office of Science, Office of High Energy Physics QuantISED Award DE-SC0019380 and by the Simons Foundation via the It From Qubit collaboration. We thank M. Van Raamsdonk, C. Waddell, D. Wakeham, M. Rozali, S. Cooper, R. Sundrum, R. Bousso, S. Kumar, Y. Wang and J. Maldacena for useful discussions.

\begin{appendices}

\appendixpage

\section{Euclidean action and Einstein-Maxwell equations}

\label{euclideansection}

The total Euclidean action for the bulk spacetime is given by
\begin{equation}
I_{bulk}=-\frac{1}{16\pi G}\int_{\mathcal{M}}d^{d+1}x\sqrt{g}(R-2\Lambda-4\pi G F_{\mu\nu}F^{\mu\nu})+I_{GHY}+I_{bound}^{em}
\label{bulkaction}
\end{equation}
where $\mathcal{M}$ is the ($d+1$)-dimensional spacetime manifold, $g$ is the determinant of the bulk metric, $R$ is the Ricci scalar, $\Lambda=-d(d-1)/(2L^2_{AdS})$ is the cosmological constant, $I_{GHY}$ is the Gibbons-Hawking-York term for the asymptotic boundary and
\begin{equation}
I_{bound}^{em}=-\int_{\partial \mathcal{M}_\infty} d^dx\sqrt{\gamma}F^{\mu\nu}\tilde{n}_\mu A_\nu
\label{bound}
\end{equation}
with $\gamma$ determinant of the metric induced on the asymptotic boundary $\partial \mathcal{M}_\infty$ and $\tilde{n}_{\mu}$ dual vector normal to the boundary. The term $I_{bound}^{em}$ is needed\cite{hawkingboundary,1-,phase} because we keep the charge fixed instead of the potential when we vary the action (\ref{bulkaction}). This choice is preferable since we are interested in controlling the charge in order to approach the extremal black hole case and obtain a sensible Euclidean solution for a near-critical brane. Additionally, the dual (pure) CFT state, will be a fixed charge state. The electromagnetic potential (in Lorentzian signature, $t$ Lorentzian time\footnote{In Euclidean signature the electromagnetic potential differs by a $-i$ factor: $A_{\tau}=-iA_t$ with $\tau$ Euclidean time. Note that all the quantities of our interest in the following involve the contraction of two $A_\tau$ with the 00-component of the Euclidean metric (which is defined with a $-$ sign with respect to the Lorentzian one). Therefore, the Lorentzian and Euclidean versions of such quantities take exactly the same form.}) $A_\mu=(A_t,\vec{0})$ for a point charge in the origin reads:
\begin{equation}
A_t=\sqrt{\frac{d-1}{8\pi G(d-2)}}\left(\frac{Q}{r^{d-2}}-\frac{Q}{r_+^{d-2}}\right)
\label{pot}
\end{equation}
where we chose the outer (event) horizon of the black hole $r=r_+$ to be the zero-potential surface and $Q$ is the charge parameter of the black hole, related to the point charge $\tilde{Q}$ (charge of the black hole) by\cite{dias,1-,phase}
\begin{equation}
Q=\sqrt{\frac{8\pi G}{(d-1)(d-2)}}\frac{\tilde{Q}}{V_{d-1}}
\label{chpar}
\end{equation}
with $V_{d-1}$ volume of the ($d-1$)-dimensional unit sphere. The electromagnetic tensor is standardly defined as $F_{\mu\nu}=\partial_\mu A_\nu-\partial_\nu A_\mu$. The action for the End-of-the-World brane is reported in equation (\ref{braneaction}), where $I_{ETW}^{em}$ has the same form (\ref{bound}) of the electromagnetic boundary term for the asymptotic boundary, with $\gamma\to h$ and $n_\mu$ dual vector normal to the ETW brane. The extrinsic curvature appearing there is defined as
\begin{equation}
K_{ab}=\nabla_\mu n_{\nu}\textrm{e}^\mu_a\textrm{e}^\nu_b,\hspace{2cm} \textrm{e}^\mu_a=\frac{dx^\mu}{dy^a}.
\label{extrinsic}
\end{equation}
A variation of the total action for the bulk and the brane leads to a set of three equations: equation (\ref{brane}) (describing the motion of the brane in the bulk spacetime and obtained by imposing Neumann boundary conditions at the brane, according to the AdS/BCFT prescription\cite{bound2,bound3}), the Einstein-Maxwell equations for the bulk
\begin{equation}
R_{\mu\nu}-\frac{1}{2}g_{\mu\nu}(R-2\Lambda)=8\pi G T_{\mu\nu}^{bulk}
\label{ein}
\end{equation}
with $R_{\mu\nu}$ Ricci tensor, and the Maxwell equation
\begin{equation}
\nabla_\mu F^{\mu\nu}=0.
\end{equation} 
The bulk stress-energy tensor appearing in equation (\ref{ein}) is the electromagnetic stress-energy tensor:
\begin{equation}
T_{\mu\nu}^{bulk}=g^{\rho\sigma}F_{\mu\rho}F_{\nu\sigma}-\frac{1}{4}g_{\mu\nu}F_{\rho\sigma}F^{\rho\sigma}.
\label{stress}
\end{equation}
The static and spherically symmetric metric reported in equation (\ref{linel}) is a solution of the Einstein-Maxwell equations (\ref{ein}) and the mass parameter $\mu$ appearing in equation (\ref{four}) (which is valid only for $d>2$) is related to the ADM mass of the black hole by
\begin{equation}
\mu=\frac{8\pi G M}{(d-1)V_{d-1}}.
\end{equation}
In order to approach, in our numerical analysis, the extremal black hole case for a fixed event horizon size by controlling the Cauchy horizon radius it is useful to express the mass and the charge parameters in terms of $r_+$ and $r_-$:
\begin{align}
&\mu=\frac{L^2_{AdS}\left[r_+^{2(d-2)}-r_-^{2(d-2)}\right]+r_+^{2d-2}-r_-^{2d-2}}{2L^2_{AdS}(r_+^{d-2}-r_-^{d-2})}; \label{massd3}\\[15pt]
&Q^2=r_+^{d-2}r_-^{d-2}\left[\frac{L^2_{AdS}\left(r_+^{d-2}-r_-^{d-2}\right)+r_+^d-r_-^d}{L^2_{AdS}\left(r_+^{d-2}-r_-^{d-2}\right)}\right].\label{charged3}
\end{align}
It is also convenient to eliminate the mass parameter from the expression of the $\tau\tau$ component of the metric $f(r)$:
\begin{equation}
f(r)=1+\frac{r^2}{L^2_{AdS}}-\frac{r_+^{d-2}}{r^{d-2}}\left(1+\frac{r_+^2}{L^2_{AdS}}\right)+\frac{Q^2}{r^{d-2}}\left(\frac{1}{r^{d-2}}-\frac{1}{r_+^{d-2}}\right)
\label{00charge}
\end{equation}
which can be rewritten, using the adimensional coordinates introduced in Section \ref{gravloc}, in a form more functional for the gravity localization analysis
\begin{equation}
f(y)=\frac{\left[y^4+\left(\gamma^2+1\right)y^2-\gamma^2q^2\right]\left(y^2-1\right)}{\gamma^2y^4}.
\label{adimmetric}
\end{equation}
For completeness, we report here also the form of the $\tau\tau$ component of the metric for the BTZ charged black hole (i.e. for $d=2$) as a function of the ADM mass $M$ and the charge of the black hole $\tilde{Q}$, in units such that $G_3=1/8$\cite{btz}:
\begin{equation}
f(r)=\frac{r^2}{L_{AdS}^2}-M-\tilde{Q}^2\ln\left(\frac{r}{L_{AdS}}\right).
\end{equation}
Finally, the Ricci scalar for a spherically symmetric metric of the form (3) takes in general the form
\begin{equation}
R=-\frac{r^2f''(r)+2r(d-1)f'(r)+(d-1)(d-2)f(r)-(d-1)(d-2)}{r^2}.
\label{ricciscalar}
\end{equation} 
Substituting equation (\ref{four}), we obtain, for $d>2$
\begin{equation}
R=-\frac{d(d+1)}{L^2_{AdS}}-\frac{(d-3)(d-2)Q^2}{r^{2d-2}}.
\label{scalar2}
\end{equation}
Note that we made use of the AdS/BCFT prescription first introduced by Takayanagi\cite{bound2}, which is suitable for ETW branes, to define our action principle. In such a formulation, the variation of the action leads to Einstein equations for the bulk without a delta contribution proportional to the tension of the brane on the right-hand side. The equations of motion for the brane are then obtained directly from the action principle (by imposing Neumann boundary conditions at the brane), instead of the Israeli junction conditions (the latter approach is employed, for instance, in ref\cite{padilla1}). Starting from this well-established formulation, the Ricci scalar has no contribution at the position of the brane proportional to the brane tension.

\section{Euclidean analysis - details}

\label{appendixb}

In order to avoid a conical singularity, the Euclidean time $\tau$ must be periodic with periodicity given by
\begin{equation}
\beta=\frac{2\pi}{k_g}.
\label{periodicity}
\end{equation}
$k_g$ is the surface gravity, which for the static and spherically symmetric black holes reads
\begin{equation}
k_g=\left.\frac{f'(r)}{2}\right|_{r=r_H}
\end{equation}
where $r_H$ is the event horizon radius. For a Reissner-Nordstr\"om black hole $r_H=r_+$. Using equation (\ref{00charge}), for $d>2$ we find the Euclidean periodicity
\begin{equation}
\beta=\frac{4\pi L^2_{AdS}r_+^{2d-3}}{(d-2)L^2_{AdS}\left[r_+^{2(d-2)}-Q^2\right]+dr_+^{2d-2}}.
\label{beta}
\end{equation}
The spherically symmetric ETW brane can be parametrize with $r=r(\tau)$. Analogously to the AdS-Schwarzschild case\cite{bhmicrostate}, the one-form dual to the unit vector normal to the brane is given by
\begin{equation}
n_\mu=N(-r',1,\mathbf{0}),\hspace{2cm} N=\sqrt{\frac{f(r)}{f^2(r)+(r')^2}}
\label{normalvector}
\end{equation}
with $r'=dr/d\tau$. The metric induced on the brane reads
\begin{equation}
\begin{aligned}
&h_{\tau\tau}=f(r)+\frac{(r')^2}{f(r)};\\
&h_{\phi_i\phi_i}=g_{\phi_i\phi_i} \hspace{1cm} i=1,..,d-1
\end{aligned}
\end{equation}
where $\phi_i$ are the coordinates on the sphere directions. Using $e^\mu_\tau=(1,r',0)$, the definition of the extrinsic curvature (\ref{extrinsic}) and equation (\ref{normalvector}), the components of the extrinsic curvature read:
\begin{equation}
    K_{\tau\tau}=-e^\mu_\tau n_\nu\nabla_\mu e^\nu_\tau=N\left[\frac{3f'(r)}{2f(r)}(r')^2-r''+\frac{f(r)f'(r)}{2}\right];
    \label{kappatautau}
\end{equation}
\begin{equation}
    K_{\phi_i\phi_i}=\frac{N f(r)}{r}h_{\phi_i\phi_i}
    \label{kappaii}
\end{equation}
where $r'=dr/d\tau$ and $r''=d^2r/d\tau^2$. Then the $\tau\tau$ component of equation (\ref{brane}) leads to equation (\ref{traj}), that we report here:
\begin{equation}
\frac{dr}{d\tau}=\pm\frac{f(r)}{Tr}\sqrt{f(r)-T^2r^2}.
\label{tautau}
\end{equation}
The $+$ ($-$) sign corresponds to the contracting (expanding) phase. Indeed, as it is clear from Figure \ref{euclidean}, for $T>0$ during the contraction the Euclidean time decreases (clockwise) from $\tau=-\tau_0$ to $\tau=-\beta/2$, therefore the $dr/d\tau>0$; during the expansion the time decreases from $\tau=\beta/2$ to $\tau=\tau_0$, and $dr/d\tau<0$. Note that choosing $T<0$ corresponds to picking the opposite normal vector to the brane, i.e. to retaining the red-shaded part of the geometry in Figure \ref{euclidean}\cite{bound5,bhmicrostate}. More precisely, since the portion of Euclidean boundary retained in our geometry is on the right, choosing $T<0$ means to consider a solution analogous to the one in Figure \ref{euclidean}, but reflected with respect to a vertical line passing through the centre of the Euclidean circle (i.e. the $\tau=\pm\beta/4$ line). The initial condition will now be $r(\tau=0)=r_0$. Note that the signs in equation (\ref{tautau}) corresponding to the expanding and contracting phase remain the same, because we did not include the absolute value for the tension. Naturally, the preparation time will be defined differently for $T<0$: $\tau_0=\Delta\tau$\cite{bound5}. This reasoning can also be understood in terms of the geometrical interpretation of the tension. Indeed, let $u_\mu=C(-f(r),-r',\mathbf{0})$ with $C=[f(r)(f^2(r)+(r')^2)]^{-1/2}$ be the unit vector tangent to the brane and $v_\mu=(0,-1/\sqrt{f(r)},\mathbf{0})$ the inward-directed unit vector normal to the asymptotic boundary. Then $\cos\theta\equiv u\cdot v\sim\sqrt{1-T^2}$ close to the boundary ($r\gg L_{AdS},r_+$), where $\theta$ is the angle between $u$ and $v$. Therefore, $T=\sin\theta$: the tension gives the angle of incidence of the brane with the asymptotic boundary\cite{bhmicrostate,bound3}. Choosing opposite values of the tension means having opposite angles of incidence of the brane with the boundary, i.e. choosing one solution or its counterpart symmetric with respect to the $\tau=\pm\beta/4$ axis. As we pointed out, in this paper we only focus our attention on the positive tension case.

It is easy to show that, if equation (\ref{tautau}) is satisfied, the other components of equation (\ref{brane}) are identically fulfilled. Indeed, using equations (\ref{kappaii}) and (\ref{tautau}), the $\phi_i\phi_i$ components of equation (\ref{brane}) lead to
\begin{equation}
     r''(t)=f(r)\left[\frac{3}{2}\frac{f(r)f'(r)}{T^2r^2}-f'(r)-\frac{f^2(r)}{T^2r^3}\right].
    \label{acc}
\end{equation}
But this equation can also be obtained by deriving equation (\ref{tautau}) and substituting again $r'$ with the right hand side of equation (\ref{tautau}). Therefore, if equation (\ref{tautau}) holds, the acceleration equation (\ref{acc}) is automatically fulfilled.

The minimum radius $r_0$ is defined by the largest zero of the RHS of equation (\ref{tautau}), i.e. by the largest solution of the equation
\begin{equation}
    f(r)=T^2r^2,
    \label{ref1}
\end{equation}
as discussed in Section \ref{euclsec}. For a vanishing tension, clearly $r_0=r_+$. Indeed, the two solutions of equation (\ref{ref1}) are $r_0^-=r_-$ and $r_0^+=r_+$, because $f(r_+)=f(r_-)=0$ by definition of horizon. Since $f(r)>0$ only when $r<r_-$ or $r>r_+$, for $0<T<1/L_{AdS}$ equation (\ref{ref1}) can have solutions only for $r<r_-$ or $r>r_+$. In particular, we note that $f(0)=\infty$ and $f(r_-)=0$, meaning that there is always a solution for $r<r_-$, which we label $r_0^-$. Additionally, given that $f(r_+)=0$ and $f(r)\sim r^2/L^2_{AdS}$ for $r\to\infty$, the condition $T<1/L_{AdS}$ guarantees that there is always a solution of equation (\ref{ref1}) for $r>r_+$ as well, that we label $r_0^+$. The latter is the largest solution, and is therefore the minimum radius of the brane $r_0$. We remark that, in the Euclidean picture, the brane always stays outside the event horizon.

A numerical evaluation shows that the ratio $r_0/r_+$ increases when the tension increases. In particular, if the black hole is big (meaning $r_+\gg L_{AdS}$), it is possible to obtain a very large ratio $r_0/r_+$ by approaching the critical value of the tension (see Figure \ref{fig3} panel (b)). If the black hole is small, the ratio is never large, preventing gravity localization on the brane.
\begin{figure}[h]
\centering
\includegraphics[height=4.5cm]{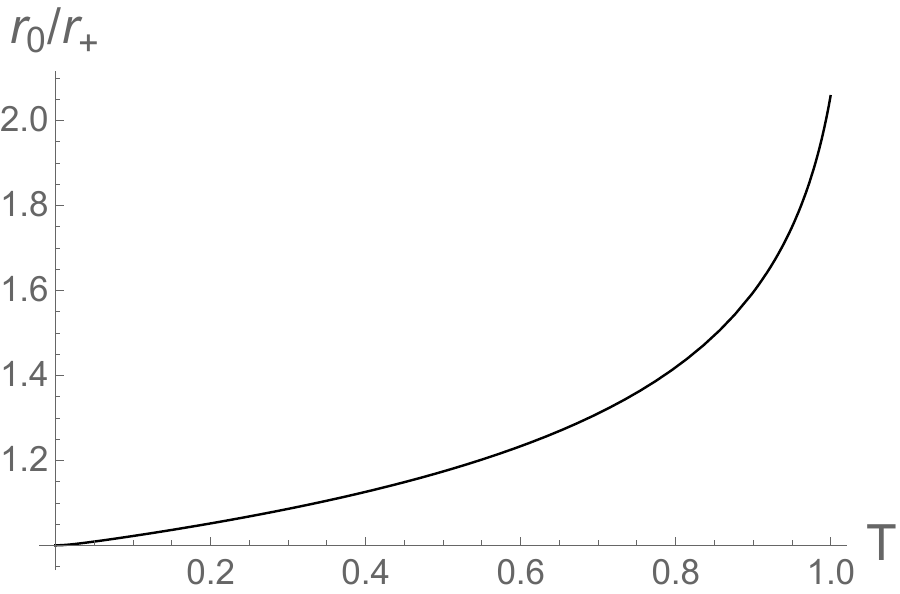}
\caption{\textbf{Minimum brane radius - Small black hole.} The minimum radius of the brane $r_0$ grows with the tension $T$, but it never becomes much larger than the black hole horizon. $d=4$, $r_+=1$, $r_-=0.99$, $L_{AdS}=1$.}
\end{figure}
Since we are interested in time-symmetric solutions with a positive-tension brane, we can require the brane to reach its minimum radius for $\tau=\pm\beta/2$, defining $r_0=r(\tau=\pm\beta/2)$. Up to a constant $\beta/2$ (which cancels out when adding up the Euclidean times necessary for the expanding and the contracting phases), the brane locus is given by
\begin{equation}
\tau(r)=\int_{r_0}^rd\hat{r}\frac{T\hat{r}}{f(\hat{r})\sqrt{f(\hat{r})-T^2\hat{r}^2}}.
\label{taur}
\end{equation}
The total Euclidean time necessary for the brane to go from $r=r_0$ to $r=\infty$, i.e. to cover half of its trajectory, is
\begin{equation}
\Delta\tau=\int_{r_0}^\infty dr \frac{Tr}{f(r)\sqrt{f(r)-T^2r^2}}.
\label{deltatau}
\end{equation}
The Euclidean preparation time $\tau_0$ is given by the residual Euclidean periodicity:
\begin{equation}
\tau_0=\frac{\beta-2\Delta\tau}{2}=\frac{2\pi L^2_{AdS}r_+^{2d-3}}{(d-2)L^2_{AdS}\left(r_+^{2(d-2)}-Q^2\right)+dr_+^{2d-2}}-\Delta\tau.
\end{equation}
For a critical brane we obtain:
\begin{equation}
\Delta\tau_{crit}=\frac{1}{L_{AdS}}\int_{r_0}^\infty dr \frac{r}{\left(1+\frac{r^2}{L^2_{AdS}}-\frac{2\mu}{r^{d-2}}+\frac{Q^2}{r^{2(d-2)}}\right)\sqrt{1-\frac{2\mu}{r^{d-2}}+\frac{Q^2}{r^{2(d-2)}}}}
\end{equation}
and this expression is logarithmically divergent at infinity. For $T>T_{crit}$, the square root $\sqrt{f(r)-T^2r^2}$ becomes imaginary for large values of $r$. Thus, $\Delta\tau$, and therefore the preparation time $\tau_0$, are well defined only for $T<T_{crit}$. 

Our numerical evaluation shows that, given a value for the outer horizon radius $r_+$ and for the brane tension $T<T_{crit}$, it is always possible to obtain a positive preparation time (i.e. a sensible Eucliden solution, and a dual description in terms of an Euclidean-time-evolved charged boundary state of a CFT living on the boundary of the right asymptotic region of the black hole) by increasing the size of the inner horizon radius $r_-$, which means by increasing the charge of the black hole. For $T\to T_{crit}$, we need $r_-\to r_+$ in order to retain a sensible Euclidean solution, i.e. the black hole must approach extremality. This feature guarantees that, for a large black hole sufficiently close to extremality, we can find a sensible Euclidean solution with the minimum radius of the brane (which is the maximum radius in Lorentzian signature) consistently larger than the black hole event horizon. As we have pointed out, this is a necessary condition in order to achieve gravity localization on the brane. In the AdS-Schwarzschild case, which is the $Q\to 0$ limit of our analysis, this result is clearly impossible to obtain. Therefore, for an uncharged black hole gravity localization and positive preparation time mututally exclude each other, and a holographic braneworld cosmology picture is not feasible.\cite{bhmicrostate}
\begin{figure}[H]
\centering
\resizebox{\textwidth}{!}{
    \begin{tabular}{ccc}
\includegraphics[height=4cm]{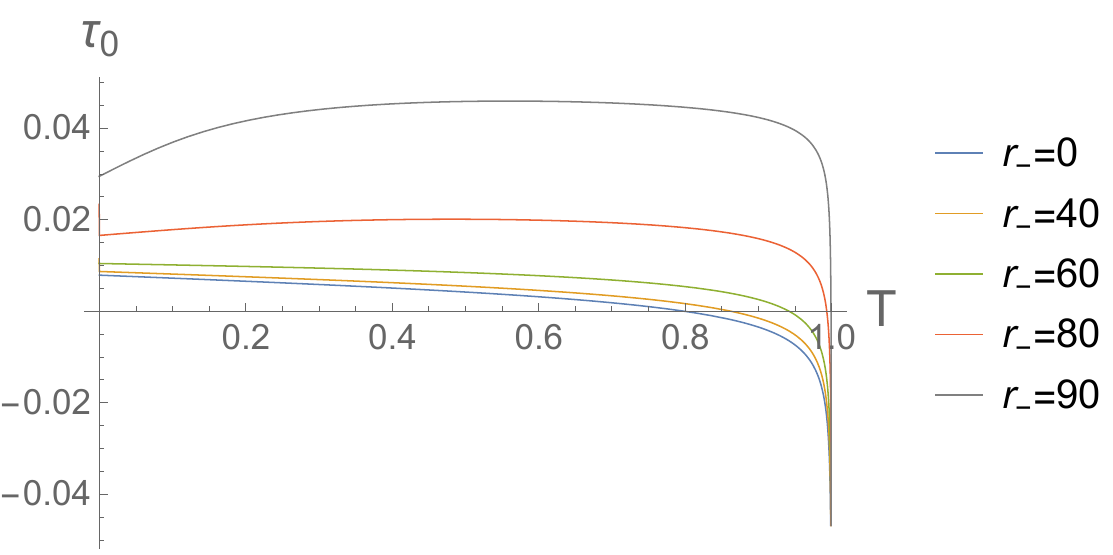}

    &   \includegraphics[height=4cm]{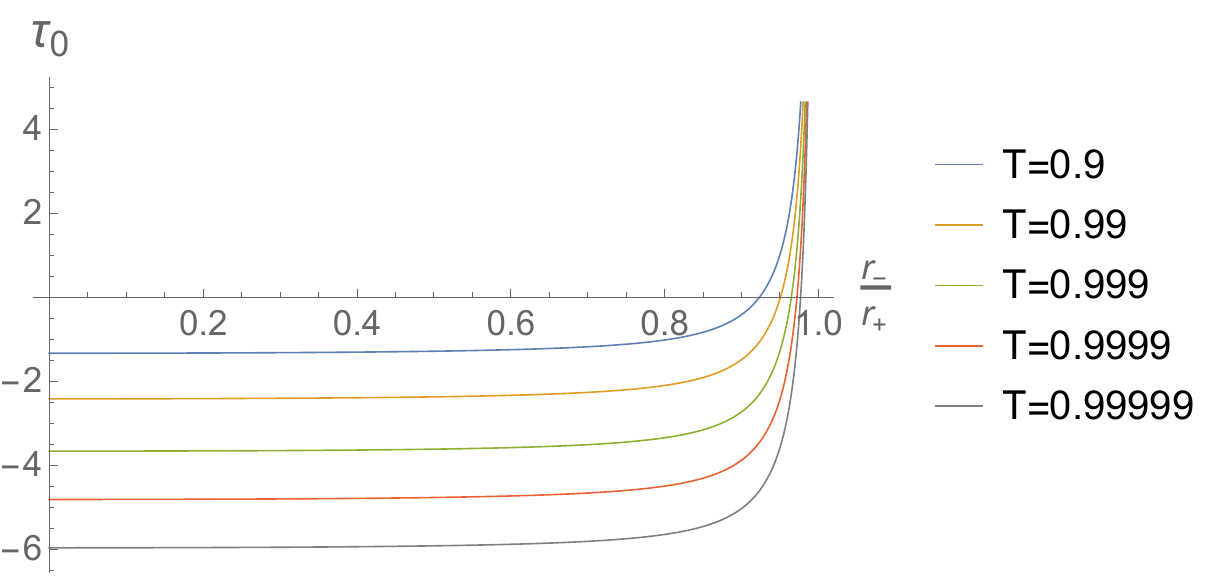}
         &   \includegraphics[height=4cm]{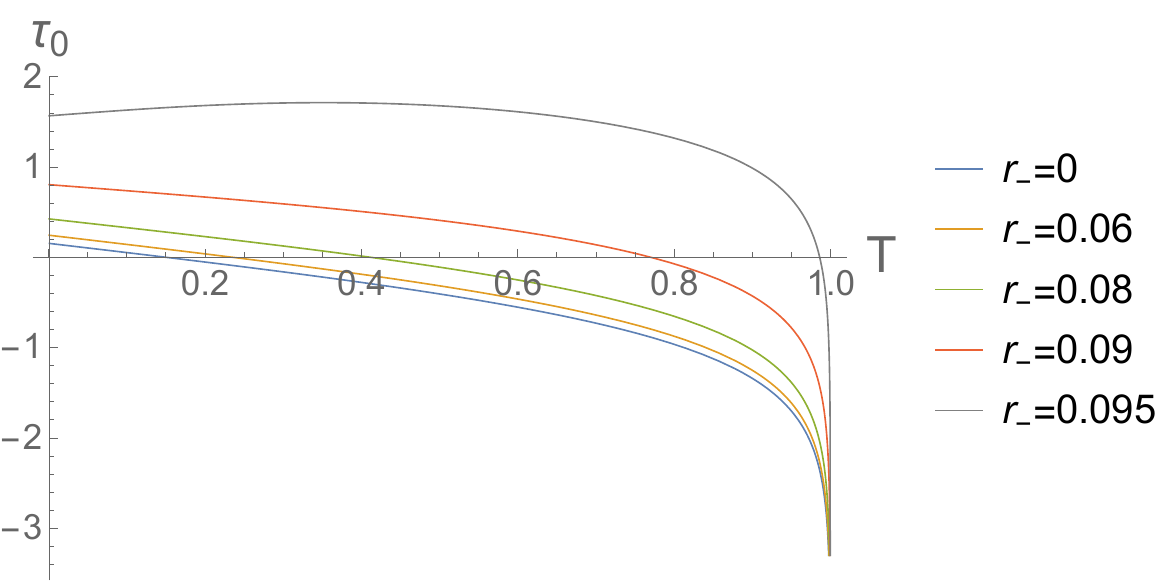}\\
         (a) & (b) &(c)
   \end{tabular}
}
\caption{\textbf{Preparation time.} (a) Large black hole case. Approaching the extremal black hole ($r_-$, and therefore the charge, increasing from bottom to top), $\tau_0$ is positive for a larger range of values for the tension $T$. $d=4$, $r_+=100$, $L_{AdS}=1$. (b)-(c) Small black hole case. The behaviour is similar to the large black hole case (the analogous of (b) is reported in Figure \ref{fig3}(a)). $d=4$, $r_+=0.1$, $L_{AdS}=1$.}
\end{figure}

\section{Action comparison}
\label{actcomp}

In order to understand if the non-extremal AdS-RN solution we studied is dominant in the Euclidean path integral, we must compare its on-shell action with other possible phases contributing to the path integral. The dominant phase will be the one with smallest action. As we have already pointed out, we will focus our attention on the fixed charge case. This choice corresponds to a canonical ensemble, where the temperature and the charge are held fixed. Note that this implies that (differently from the fixed potential case) the variation of the potential does not vanish on the boundary. For this reason, we need to add the electromagnetic boundary terms (\ref{bound}) for the asymptotic boundary and the brane to the action. For fixed charge, the other possible phase is represented by the extremal AdS-RN black hole with the same charge as the corresponding non-extremal solution (for a complete treatment of the AdS-RN phase structure for fixed charge and fixed potential without the ETW brane see refs.\cite{1-,phase}). The CFT state corresponding to the fixed charge ensemble is the Euclidean-time-evolved boundary state with fixed charge reported in equation (\ref{state}). It is clear that the two Euclidean solutions contributing to the gravity path integral dual to the state (\ref{state}) must have the same charge and the same preparation time $\tau_0$. We will use this property in order to match the two geometries in the asymptotic region, procedure needed in order to regularize the Euclidean action and compare the two actions.\cite{1-,phase,bhmicrostate} We will now briefly review some useful properties of the extremal AdS-RN black hole.

\subsection*{Extremal AdS-RN black hole}

For the extremal AdS-Reissner-Nordstr\"om black hole, the two solutions $r_+$ and $r_-$ of $f_e(r)=0$ coincide, defining the extremal horizon radius $r_e$. Therefore, also the relationship $\left.f'_e(r)\right|_{r=r_e}=0$ must hold. By requiring the two conditions to be satisfied, we obtain for the extremal mass and charge parameters:
\begin{align}
&2\mu_e=2r_e^{d-2}+\frac{2d-2}{d-2}\frac{r_e^d}{L^2_{AdS}};\label{extm}\\
&Q^2_e=r_e^{2d-4}+\frac{d}{d-2}\frac{r_e^{2d-2}}{L^2_{AdS}}.\label{extch}
\end{align}
We remark how, by fixing the charge of the black hole, the extremal horizon radius is uniquely determined.

Another important feature of the extremal black hole is that, since $\left.f'_e(r)\right|_{r=r_e}=0$, the Euclidean periodicity (\ref{periodicity}) diverges, i.e. the temperature of the extremal black hole vanishes. Nonetheless, it has been demonstrated \cite{1-,phase,tempestr} that an arbitrary Euclidean periodicity can be chosen for the extremal RN black hole without running into a conical singularity. Physically, this means that an extremal RN black hole can be in thermodynamic equilibrium with a thermal bath at an arbitrary temperature. In the following we will use this feature, fixing a total Euclidean periodicity $\beta_e$ for the extremal black hole that will be determined by matching the geometries of the non-extremal and the extremal black holes in the asymptotic regions. Given $\beta_e$, the extremal horizon radius $r_e$ and the expression (\ref{extch}) for the extremal charge, the same Euclidean analysis carried out in Appendix \ref{euclideansection} can be applied to the extremal case.

Before explicitly evaluating the difference of the two Euclidean actions for the non-extremal and extremal black holes, one more remark is needed. In general, the phase structure is more complicated than the one we are going to study. In particular, for a given small charge $Q<Q_{crit}$ and some range of temperatures, three different non-extremal black holes with the same temperature but different horizon radius can exist \cite{1-}. Indeed, for $Q<Q_{crit}$, there exist turning points for the Euclidean periodicity $\beta$ as a function of the external horizon $r_+$, which disappear for $Q>Q_{crit}$. The critical charge can be obtained from the condition $\partial_{r_+}\beta=0=\partial_{r_+}^2\beta$, and reads \cite{1-}:
\begin{equation}
Q^2_{crit}=\frac{1}{(d-1)(2d-3)}\left[\frac{(d-2)^2}{d(d-1)}\right]L_{AdS}^{2d-4}.
\end{equation}
The coexistence of different non-extremal black holes implies the necessity of studying which one of them has the smallest action, before a comparison with the extremal case can be made. To do so, we should fix an Euclidean periodicity $\beta$ and a charge $Q$, find the different corresponding values of $r_+$ and compare the actions for each of them.

Nonetheless, since, as we have already pointed out, we are interested in solutions involving near-critical branes, and this requires to have a large black hole near extremality in order to have a positive preparation time, we are safely into the region where only one non-extremal phase exists \cite{1-}. Note that the parameter we want to fix is the preparation time $\tau_0$, which is meaningful in the dual CFT, and not the Euclidean periodicity $\beta$. But for $Q>Q_{crit}$, a numerical analysis shows that, given a tension $T$, when $\tau_0>0$ it is in a one-to-one correspondence with $\beta$. In other words, fixing $Q$, $T$ and $\beta$ or fixing $Q$, $T$ and $\tau_0$ is completely equivalent for the region of parameters of our interest, and there is only one possible non-extremal phase for each choice of parameters. Therefore, we are allowed to choose $r_+$ and $r_-$ independently, and a tension $T$ (this is equivalent to choose a tension $T$, a charge $Q$ and a preparation time $\tau_0$) and compare directly the resulting action with the corresponding extremal action with the same charge.

\subsection*{Bulk action}

Let us evaluate the three terms of the bulk action (\ref{bulkaction}) separately in the non-extremal case. First, using equation (\ref{pot})\footnote{Using equation (\ref{pot}), we must raise the indices with the Lorentzian metric, computing $F_{\mu\nu}F^{\mu\nu}$ in the Lorentzian case. As we have pointed out, this yields exactly the same result that we would get using the Euclidean version of the electromagnetic potential $A_{\tau}=-iA_t$ and raising the indices with the Euclidean metric. In other words, $F_{\mu\nu}F^{\mu\nu}$ is unchanged under Wick's rotation. The same is true for the combination $F^{\mu\nu}n_\mu A_\nu$ appearing in the electromagnetic boundary term.}, the only non vanishing components of the electromagnetic tensor are $F_{\tau r}=-F_{r\tau}$, and we obtain
\begin{equation}
F_{\mu\nu}F^{\mu\nu}=-\frac{(d-1)(d-2)}{4\pi G}\frac{Q^2}{r^{2d-2}}.
\label{ff}
\end{equation}
Using equation (\ref{ricciscalar}) for the Ricci scalar, the relationship between $\Lambda$ and $L_{AdS}$, and noting that $\sqrt{g}=r^{d-1}$, the first term of the bulk action reads
\begin{equation}
I_{bulk}^{(1)}=\frac{dV_{d-1}}{8\pi G L^2_{AdS}}\int drd\tau r^{d-1}-\frac{(d-2)V_{d-1}Q^2}{8\pi G}\int drd\tau\frac{1}{r^{d-1}}.
\end{equation}
Recalling that part of the Euclidean geometry is cut off by the ETW brane parametrized by $\tau(r)$, we can rewrite this integral as the integral in the entire spacetime minus the excised part. In order to avoid divergences, we must also introduce a cut-off $R$ in the asymptotic region. After a few steps, the result turns out to be
\begin{equation}
\begin{split}
I_{bulk}^{(1)}=&\frac{V_{d-1}}{8\pi GL^2_{AdS}}\left[\beta(R^d-r_+^d)-2\left.(r^d\tau(r))\right|_{r_0}^R+2\int_{r_0}^Rdrr^d\frac{\partial\tau}{\partial r}\right]\\[15pt]
&+\frac{V_{d-1}Q^2}{8\pi G}\left[\beta\left(\frac{1}{R^{d-2}}-\frac{1}{r_+^{d-2}}\right)-2\left.\left(\frac{\tau(r)}{r^{d-2}}\right)\right|_{r_0}^R+2\int_{r_0}^Rdr\frac{1}{r^{d-2}}\frac{\partial\tau}{\partial r}\right]
\end{split}
\label{totbulk}
\end{equation}
where $\beta$ is given by equation (\ref{beta}).

The Gibbons-Hawking-York term $I_{GHY}$ gives a vanishing contribution to the action difference for asymptotically AdS spacetimes\cite{1-,phase,counterterm} and therefore we can neglect it. The electromagnetic boundary term finally reads
\begin{equation}
I_{bound}^{em}=-\frac{(d-1)V_{d-1}Q^2\tau_0}{4\pi G}\left(\frac{1}{R^{d-2}}-\frac{1}{r_+^{d-2}}\right).
\label{expbound}
\end{equation}

\subsection*{Brane action}

Tracing equation (\ref{brane}) we obtain
\begin{equation}
K-(d-1)T=T.
\end{equation}
Parametrizing again the brane with $\tau(r)$, and using equation (\ref{tautau}), the determinant of the metric induced on the brane reads:
\begin{equation}
\sqrt{h}=\pm\frac{1}{T}f(r)r^{d-2}\frac{\partial\tau}{\partial r}
\label{detmetricbrane}
\end{equation}
where the $+$ ($-$) sign is for the contracting (expanding) phase. Considering the contributions of both these two phases and using the expression (\ref{normalvector}) for the dual vector normal to the brane, the total action for the ETW brane reads
\begin{equation}
\begin{split}
I_{ETW}=&-\frac{V_{d-1}}{4\pi G}\int_{r_0}^Rdrf(r)r^{d-2}\frac{\partial\tau}{\partial r}\\[10pt]
&+\frac{(d-1)V_{d-1}Q^2}{4\pi G r_+^{d-2}}\tau(r)\Big|_{r_0}^R-\frac{(d-1)V_{d-1}Q^2}{4\pi G}\int_{r_0}^Rdr\frac{1}{r^{d-2}}\frac{d\tau}{dr}.
\end{split}
\label{etwem}
\end{equation}

\subsection*{Action difference}

Using the definition (\ref{taur}), we note that $\tau(r_0)=0$. Additionally, $\lim_{R\to\infty}\tau(R)=\Delta\tau$ and $\beta-2\Delta\tau=2\tau_0$. Since, after subtracting the extremal action from the non-extremal one, we will take the limit $R\to\infty$, for simplicity of notation we will substitute now $\tau(R)\to\Delta\tau$ and $\beta-2\tau(R)\to 2\tau_0$, even if the limit has not been taken yet. Using the above cited relationships, the total action for the non-extremal black hole can be cast in the form
\begin{equation}
\begin{split}
I_{tot}=&\frac{V_{d-1}}{8\pi GL^2_{AdS}}\left\{2\tau_0R^d-\beta r_+^d-\frac{2(d-2)L^2_{AdS}Q^2}{R^{d-2}}\tau_0+\frac{(d-2)L^2_{AdS}Q^2}{r_+^{d-2}}\beta\right.\\
&\left.+2\int_{r_0}^Rdr\left[\left(r^d-\frac{(d-2)L^2_{AdS}Q^2}{r^{d-2}}-L^2_{AdS}r^{d-2}f(r)\right)\frac{\partial\tau}{\partial r}\right]\right\}.
\end{split}
\label{act}
\end{equation}
We must now subtract from this action the equivalent action for the extremal black hole after matching the geometries in the asymptotic region, and then take the limit $R\to\infty$. The total action for the extremal black hole is given by equation (\ref{act}) where all the quantities are substituted by their extremal counterparts ($r_+\to r_e$, $\tau_0\to\tau_0^e$, $\beta\to\beta_e$, $r_0\to r_0^e$, $f(r)\to f_e(r)$, $Q\to Q_e$). As we have already mentioned, the Euclidean analysis carried out for the non extremal black hole is still valid by substituting $r_+\to r_e$, and the total Euclidean periodicity $\beta_e$ can be chosen arbitrarily \cite{tempestr}. Since we are interested in the fixed charge case, we can choose a charge, which will be the same for both the black holes ($Q=Q_e$), using equation (\ref{charged3}) and selecting values for the outer and inner horizon radii ($r_+$ and $r_-$) of the non-extremal black hole. The extremal horizon radius $r_e$ is then uniquely determined by equation (\ref{extch}). Therefore $r_e$ can be written as a function of $r_+$ and $r_-$ only. We will use this feature in our numerical analysis.

The choice of the (arbitrary) Euclidean periodicity for the extremal black hole is determined by the matching of the two geometries in the asymptotic region. In particular, we must match the proper Euclidean preparation times in the asymptotic region (i.e. for $r=R$):
\begin{equation}
2\tau_0\sqrt{f(R)}=2\tau^e_0\sqrt{f_e(R)}.
\label{matching}
\end{equation}
At lowest order in $1/R$, equation (\ref{matching}) gives
\begin{equation}
\tau_0^e\sim\tau_0(1+A)
\label{match}
\end{equation}
where we have defined:
\begin{equation}
A=\frac{L^2_{AdS}}{R^d}(\mu_e-\mu)=\frac{L^2_{AdS}}{R^d}\left(r_e^{d-2}+\frac{d-1}{d-2}\frac{r_e^d}{L^2_{AdS}}-\frac{r_+^{d-2}}{2}-\frac{r_+^d}{2L^2_{AdS}}-\frac{Q^2}{2r_+^{d-2}}\right).
\end{equation}
We emphasize that, at the asymptotic boundary (i.e. for $R\to\infty$), $\tau_0^e=\tau_0$, as we expect for two geometries dual to the state (\ref{state}). The total Euclidean periodicity for the extremal black hole can then be set to be
\begin{equation}
\beta_e=2\tau_0^e+2\Delta\tau_e=2\tau_0(1+A)+2\Delta\tau_e
\label{tempe}
\end{equation}
where $\Delta\tau_e$ is defined in perfect analogy with the non-extremal case. Using equation (\ref{tempe}) and the matching condition (\ref{match}) and taking the limit $R\to\infty$, after some manipulation the difference between the non-extremal and the extremal action finally takes the form
\begin{equation}
\begin{split}
\Delta I\equiv&\frac{8\pi GL^2_{AdS}}{V_{d-1}}\lim_{R\to\infty}(I-I_{ext})=\\[10pt]
&-\frac{2r_e^d\tau_0}{d-2}-2L^2_{AdS}r_e^{d-2}\tau_0+L^2_{AdS}r_+^{d-2}\tau_0+r_+^d\tau_0-\beta r_+^d+\frac{L^2_{AdS}Q^2\tau_0}{r_+^{d-2}}\\[10pt]
&+2r_e^d\Delta\tau_e+\frac{(d-2)L^2_{AdS}Q^2\beta}{r_+^{d-2}}-\frac{2(d-2)L^2_{AdS}Q^2\tau_0}{r_e^{d-2}}-\frac{2(d-2)L^2_{AdS}Q^2\Delta\tau_e}{r_e^{d-2}}\\[10pt]
&-2L^2_{AdS}\int_{r_0^e}^{r_0}dr\left\{\left[-r^{d-2}+r_e^{d-2}\left(1+\frac{r_e^2}{L^2_{AdS}}\right)+\frac{Q^2}{r_e^{d-2}}-\frac{(d-1)Q^2}{r^{d-2}}\right]\frac{Tr}{f_e(r)\sqrt{f_e(r)-T^2r^2}}\right\}\\[10pt]
&+2L^2_{AdS}\int_{r_0}^\infty dr\left\{\left[-r^{d-2}+r_+^{d-2}\left(1+\frac{r_+^2}{L^2_{AdS}}\right)+\frac{Q^2}{r_+^{d-2}}-\frac{(d-1)Q^2}{r^{d-2}}\right]\frac{Tr}{f(r)\sqrt{f(r)-T^2r^2}}\right.\\[10pt]
&\left.-
\left[-r^{d-2}+r_e^{d-2}\left(1+\frac{r_e^2}{L^2_{AdS}}\right)+\frac{Q^2}{r_e^{d-2}}-\frac{(d-1)Q^2}{r^{d-2}}\right]\frac{Tr}{f_e(r)\sqrt{f_e(r)-T^2r^2}}\right\}.
\end{split}
\label{deltai}
\end{equation}
When such a difference is negative, the non-extremal black hole phase has smaller action and is therefore dominant in the ensemble.

A numerical evaluation shows that, for $d=4$ and for a range of parameters such that the Euclidean solution for the non-extremal black hole is sensible (i.e. $\tau_0>0$), $\Delta I$ is always negative. Therefore, differently from the AdS-Schwarzschild case\cite{bhmicrostate}, when the non-extremal AdS-RN solution admits a dual CFT description, it is also always dominant in the gravity path integral.
\begin{figure}[H]
\centering
\resizebox{\textwidth}{!}{
    \begin{tabular}{cc}
\includegraphics[height=4cm]{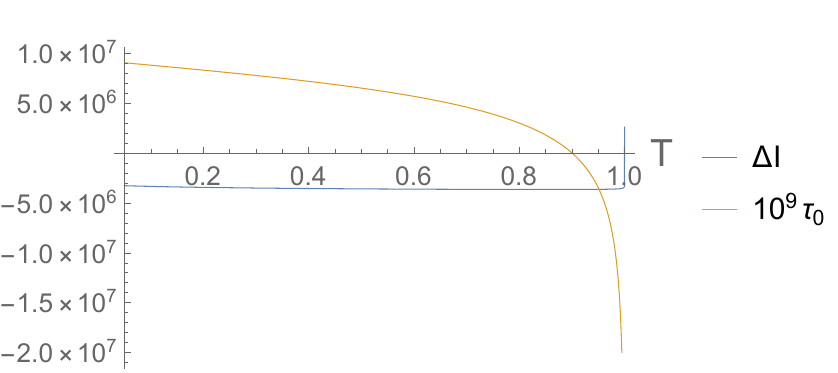}
    &   \includegraphics[height=4cm]{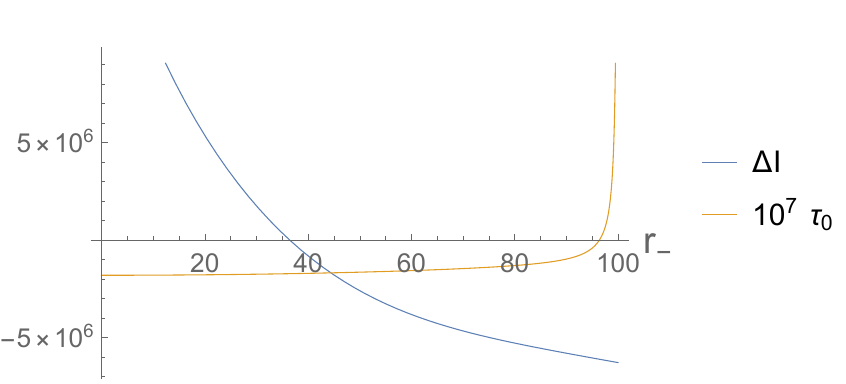}\\
         (a) & (b)
   \end{tabular}
}
\caption{\textbf{Action difference - Large black hole.} Results for four spatial dimensions ($d=4$). We set the outer horizon radius to be $r_+=100$, with the AdS length given by $L_{AdS}=1$. (a) We chose $r_-=50$ for the inner horizon radius. When the tension $T$ is not too close to its critical value $T_{crit}=1/L_{AdS}=1$, the preparation time $\tau_0$ is positive (i.e. the solution is Euclidean-sensible) and the action difference $\Delta I$ is negative, meaning that the non-extremal phase is always dominant in the path integral. (b) For a near-critical brane (we chose a value $T=0.99999\sim T_{crit}$ for the brane tension), when the black hole is sufficiently close to extremality (i.e. $r_-\to r_+$), the preparation time $\tau_0$ becomes positive, and therefore the Euclidean solution is sensible. The non-extremal phase is already dominant in the ensemble (i.e. the action difference $\Delta I$ is negative) well before it happens. The same properties hold also for the small black hole case.}
\end{figure}

\section{Lorentzian analysis and braneworld cosmology - details}

\begin{figure}[H]
\centering
\resizebox{\textwidth}{!}{
    \begin{tabular}{cc}
\includegraphics[height=4cm]{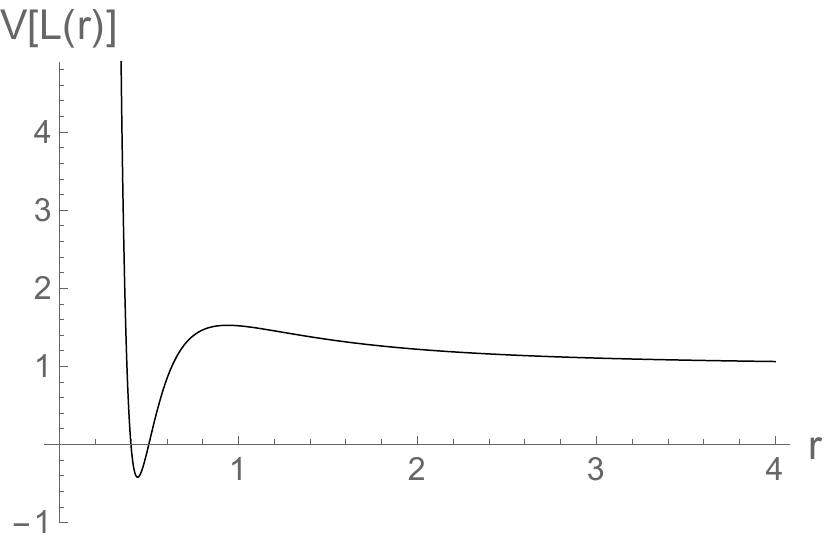}
    &   \includegraphics[height=4cm]{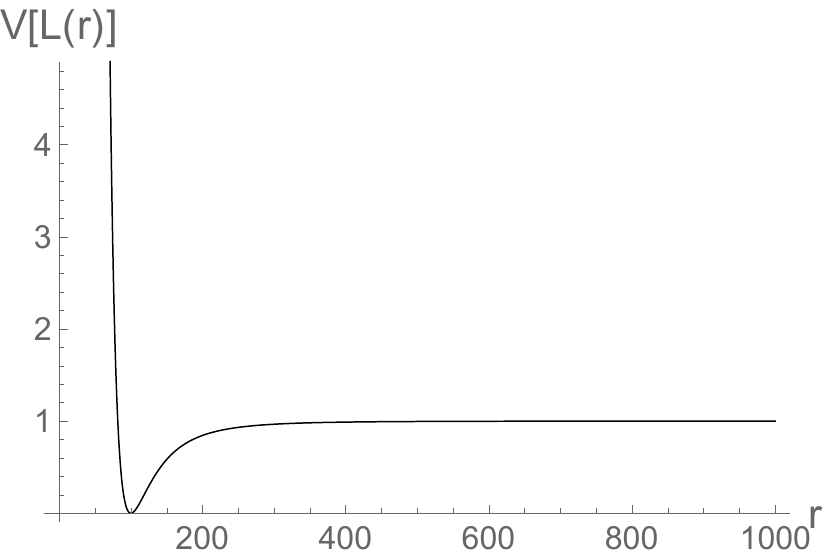}\\
         (a) & (b)
   \end{tabular}
}
\caption{\textbf{Potential V[L(r)].} (a) Small black hole case. The peak at $r=r_M$ is particularly evident. $d=4$, $r_+=0.5$, $r_-=0.4$, $L_{AdS}=1$. (b) Large black hole case. $r_M$ is very large and the value of the potential at the peak is almost indistinguishable from the asymptotic value. $d=4$, $r_+=100$, $r_-=99.9$, $L_{AdS}=1$.}
\label{lorpot}
\end{figure}

The initial condition for the real time evolution is given by the $\tau=0,\pm\beta/2$ slice of the Euclidean geometry, where the brane reaches its minimum radius $r_0$. The time coordinate is analytically continued $\tau\to -it$, with $t$ Schwarzschild time. Further time evolution will be in Lorentzian signature. The brane locus is then given by
\begin{equation}
t(r)=\int_{r_0}^rd\hat{r}\frac{T\hat{r}}{f(\hat{r})\sqrt{T^2\hat{r}^2-f(\hat{r})}}.
\label{schw}
\end{equation}
We explained that, in the Lorentzian case, the range of the radial coordinate is $r_0^-<r_0<r_0^+\equiv r_0$. Let us now focus on the Friedmann equation (considered in equation (\ref{friedmain})) describing the evolution of the brane in terms of its proper time $\lambda$. By defining $L(r)=\ln r$ and $L_+=L(r_+)$, it can be rewritten as
\begin{equation}
\dot{L}^2+V(L)=T^2
\end{equation}
where
\begin{equation}
V[L(r)]=\frac{f(r)}{r^2}.
\label{poten}
\end{equation}
In this new coordinate, the motion of the brane can be regarded as the one of a particle with energy $T^2$ in the presence of a potential $V(L)$. The potential (\ref{poten}) naturally vanishes for $r=r_+$ and $r=r_-$. It has a local maximum and a local minimum in $r_M$ and $r_m$ respectively. For the local minimum $r_-<r_m<r_+$ and $V[L(r_m)]<0$, while for the local maximum $r_M>r_+$ and $V[L(r_M)]>1/L^2_{AdS}=T_{crit}^2$. Additionally $\lim_{r\to\infty}V[L(r)]=T_{crit}^2$, while $\lim_{r\to 0}V[L(r)]=\infty$. The latter behaviour confirms that the brane radius cannot vanish, i.e. the brane undergoes a bounce for a minimum radius $r_0^-$. Differently from the small black hole case ($r_+\lesssim L_{AdS}$), in the large black hole case ($r_+\gg L_{AdS}$) the minimum of the potential is $V[L(r_m)]\sim 0$ and its local maximum is $V[L(r_M)]\sim T_{crit}^2$, the latter meaning that the value of the potential at the peak is almost indistinguishable from its asymptotic value. As an example we report in Figure \ref{lorpot} a plot of the potential $V[L(r)]$ as a function of the radius of the brane $r$ in the small and in the large black hole cases.
\noindent The brane trajectory will be determined by the value of its tension (i.e. the energy of the particle). In particular, analogously to the AdS-Schwarzschild case \cite{bhmicrostate}, we can distinguish in general five different cases:

\begin{enumerate}
\item $\mathbf{T^2>V[L(r_M)]}.$ The equation $f(r_0)=T^2r_0^2$ has only one positive solution ($r_0^-$), which is by definition the position of the brane at $\lambda=0$ (where $\lambda$ is the brane proper time). Therefore, the brane radius decreases from $r=\infty$ to $r=r_0^-$ for negative time and then increases monotonically from $r=r_0^-$ to $r=\infty$ for positive time.
\item $\mathbf{T^2=V[L(r_M)]}.$ The brane has constant radius $r=r_M$. This corresponds to an Einstein static universe.
\item $\mathbf{T^2_{crit}<T^2<V[L(r_M)]}$\textbf{ - Large $\mathbf{r}$ branch.} The equation $f(r_0)=T^2r_0^2$ admits three solutions. In particular, if we order such solutions as $r_0^{(3)}>r_0^{(2)}>r_0^{(1)}$, the quantity $\sqrt{T^2r^2-f(r)}$ is real for $r>r_0^{(3)}$ and for $r_0^{(1)}<r<r_0^{(2)}$. Thus, for large $r$, the behaviour is similar to the first case, with the brane trajectory starting from $r=\infty$, shrinking to $r=r_0^{(3)}$ for $\lambda=0$ and expanding again to $r=\infty$.
\item $\mathbf{T^2_{crit}<T^2<V[L(r_M)]}$\textbf{ - Small $\mathbf{r}$ branch.} This is the same situation as in the previous case, but with $r_0^{(1)}<r<r_0^{(2)}$. The brane expands from $r=r_0^{(1)}$ to $r=r_0^{(2)}$ at $\lambda=0$, and contracts again to $r=r_0^{(1)}$.
\item $\mathbf{T^2<T^2_{crit}}.$ In this case, as we have already pointed out, the equation $f(r_0)=T^2r_0^2$ has two solutions, $r_0^+$ and $r_0^-$. The brane expands from $r=r_0^-$, emerges from the horizon, reaches its maximum radius $r=r_0^+$ for $\lambda=0$, and shrinks again to $r=r_0^-$. This trajectory is completed in a finite amount of proper time, as we will see.
\end{enumerate}

We remark that, as we have pointed out, our Euclidean analysis is feasible only for $T<T_{crit}$, condition needed to have a well-defined preparation time $\tau_0$. Therefore only the last situation is of interest for our analysis. Nonetheless, a continuously expanding brane with $T>T_{crit}$ (first and third case) implies, from a braneworld cosmology point of view, a 4-dimensional Friedmann universe with a positive cosmological constant, which undergoes for late times a de Sitter expanding phase. This cosmological scenario is clearly more in accordance with the behaviour of our observed universe with respect to the Big Bounce cosmology of the last case. Therefore, an interesting future development of this work could be its extension to over-critical branes, which probably requires a different class of CFT states in the dual theory.

Focusing on the last case, a few final remarks are worth of attention. First, the choice of the sign of the tension $T$ determines which vector normal to the brane we are considering \cite{bound5}, i.e. which one of the two sides of the brane we are retaining. The choice $T>0$ corresponds to a Lorentzian geometry which contains the complete right asymptotic region and part of the left one, ending on the ETW brane (see Figure \ref{penrose}). Additionally, the total brane proper time needed to complete the trajectory from $r_0^-$ to $r_0^+$ and back is given by
\begin{equation}
\lambda_{tot}=2\int_{r_0^-}^{r_0^+}d\hat{r}\frac{1}{\sqrt{T^2\hat{r}^2-f(\hat{r})}}
\label{propertot}
\end{equation}
and is clearly finite. This expression is the one used in Table \ref{timescales} in order to quantify the portion of the trajectory of the brane where the adiabatic approximation is reliable. Finally, from the definition of proper time, we find
\begin{equation}
\dot{t}=\frac{dt}{d\lambda}=\pm\frac{Tr}{f(r)}.
\end{equation}
Referring again to Figure \ref{penrose}, $f(r)$ has a definite (negative) sign in region II ($r_-<r<r_+$). This feature implies that the Schwarzschild time is monotonic with respect to the proper time during the brane trajectory in that region. Thus, if in the contraction phase the brane crossed the outer horizon, for instance, from the left asymptotic region I' (i.e. it crossed the line $r=r_+$, $t=-\infty$), it must cross the inner horizon to the right internal region III (i.e. it must cross the line $r=r_-$, $t=\infty$), cutting off completely the left internal region. As far as we know, this was first noted in ref\cite{instability}. The existence of the minimum radius $r_0^-$, where the brane reverses again its motion, suggests that, gluing together more spacetime patches as in Figure 1, we can obtain a cyclical Big Bounce cosmology (with the caveat that gravity is not always localized on the brane during the trajectory). However, it is not clear if the effective semiclassical analysis we used to derive the brane trajectory is reliable when the brane is close to or inside the inner horizon. Indeed, the instability of the Cauchy horizon against even small excitations of bulk fields implies that curvature sigularities may appear\cite{instability}, preventing a smooth evolution of the brane trajectory. 
\begin{figure}[H]
\centering
\resizebox{\textwidth}{!}{
    \begin{tabular}{cc}
\includegraphics[height=4cm]{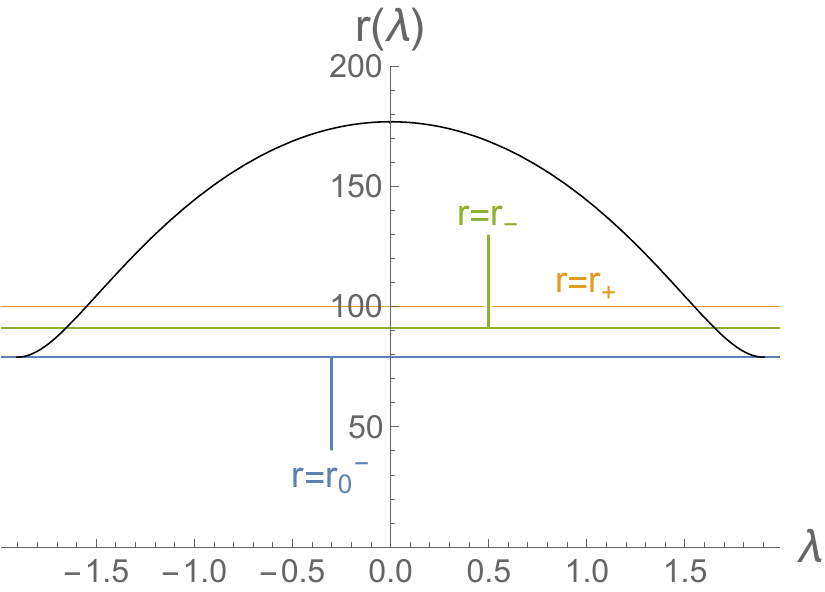}
    &   \includegraphics[height=4cm]{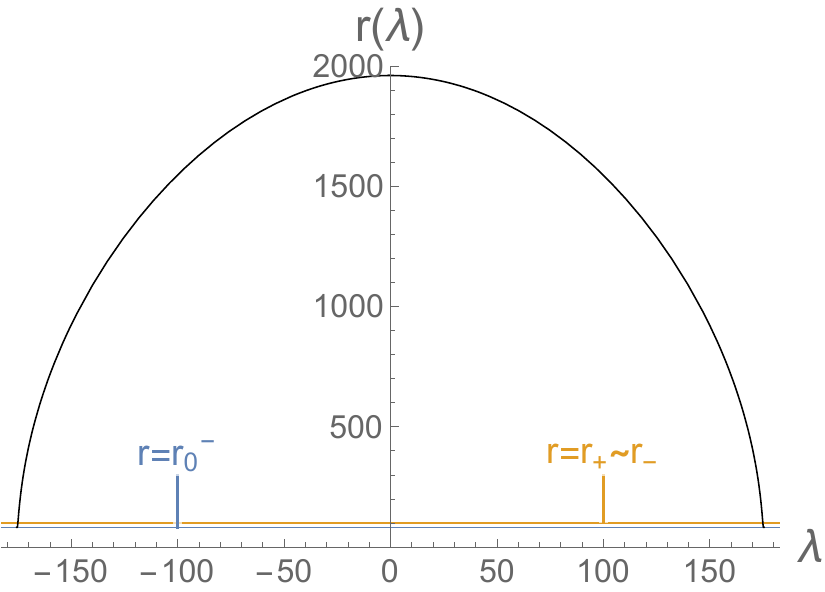}\\
         (a) & (b)
   \end{tabular}
}
\caption{\textbf{Proper brane trajectory $\mathbf{r(\lambda)}$.} Results for four spatial dimensions ($d=4$), where we chose the AdS length to be $L_{AdS}=1$. (a) Brane radius $r(\lambda)$ as a function of the brane proper time $\lambda$ for a subcritical brane of tension $T=0.89$. During its trajectory, the brane radius expands to a maximum value $r_0$ larger than the black hole outer horizon $r_+=100$, and shrinks to a minimum value $r_0^-$ smaller than the black hole inner horizon $r_-=91$. (b) For larger values of the brane tension ($T=0.99999\sim T_{crit}=1/L_{AdS}=1$) and a near-extremal black hole ($r_+=100$, $r_-=99.9$), the maximum radius of the brane $r_0$ is larger, and the inversion at the minimum radius $r_0^-$ becomes sharper. In both cases the brane trajectory resembles the one qualitatively described in Figure \ref{penrose}.}
\end{figure}

\section{Gravity Localization and trapping coefficient method}
\label{gravlocdet}

Before clarifying the technical details of the gravity localization analysis, let us qualitatively understand what conditions we expect to be necessary in order to obtain an effective 4-dimensional description of gravity on the ETW brane.

\begin{enumerate}
\item  The radius of the brane must be much larger than the event horizon size and the AdS radius: $r_b\gg r_+,L_{AdS}$. Since gravity in a 5-dimensional spacetime is unavoidably influenced by the presence of the fifth dimension, a 4-dimensional description of gravity is achievable only in a region where the gravitational effects directed along the fifth dimension and due to the presence of the black hole are weak. If the brane is too close to the black hole horizon, the gravitational attraction of the black hole is strong and a graviton localized on the brane will certainly ``leak'' in the fifth dimension, eventually falling into the black hole horizon. On the other hand, if the brane is far enough from the black hole horizon and deep in the asymptotically AdS region ($r_b\gg L_{AdS}$), we recover a setup very similar to Randall-Sundrum II (RS II) model, with the brane embedded in a region of the spacetime which is nearly AdS. Thus we can expect to recover, at least locally, gravity localized on the brane. It is somehow surprising that gravity localization is lost if the radius of the brane is too large, as we will see.
\item The time scale of the motion of the brane must be much smaller than the time scale of oscillation of a graviton: $1/H\ll t_o$. This condition is needed to treat the motion of the brane as adiabatic with respect to gravitational perturbations. In other words, under this condition we can consider the brane as effectively static during the propagation of a graviton. As we will see, this assumption turns out to be pivotal in our analysis. Indeed, in the case of a moving brane, it is non trivial to define a boundary condition for the graviton wavefunction at the position of the brane.
\item $\frac{[r'(t)]^2}{f(r)}\ll f(r)$ or equivalently $\frac{[y'(t)]^2}{f(y)}\ll f(y)$. It is clear looking at the definition of the brane proper time that this condition guarantees that the properties of the graviton are interpreted, up to a constant redshift factor $\sqrt{f(r_b)}=\sqrt{f(y_b)}$, in the same way by a bulk observer (from the perspective of which we develop our analysis) and by an observer comoving with the brane.
\end{enumerate}

We remark that, differently from the AdS-Schwarzschild case, if the black hole is large, the first condition can be satisfied for part of the brane trajectory while still retaining a positive preparation time, provided that the brane is near-critical and the black hole near-extremal. The presence of the black hole unavoidably alters the RS II model, therefore we expect gravity localization to be only a local phenomenon, and valid only for a limited range of time.

\subsection{TT perturbations and Schr\"odinger equation}
\label{appendixe1}

The first step to find the Schr\"odinger equation reported in Section \ref{gravloc} is to derive the linearized Einstein equations about the AdS-RN background. Since we are considering only the tensor component of the metric perturbation, the perturbations of the electromagnetic field (which couple only to the scalar and the vector perturbations of the metric\cite{kodama}) can be neglected. Therefore, the only terms arising from the perturbation of the electromagnetic stress-energy tensor will be the ones linear in the metric perturbation and with an unperturbed electromagnetic field. For a transverse-traceless (TT) perturbation (i.e. one that satisfies $\delta g^\mu_\mu=0=\nabla^\mu\delta g_{\mu\nu}$) the linearized Einstein equations read
\begin{equation}
\delta R_{\mu\nu}=\left(\frac{2}{d-1}\Lambda+\frac{(d-2)Q^2}{r^{2d-2}}\right)\delta g_{\mu\nu}
\label{pert}
\end{equation}
with
\begin{equation}
\delta R_{\mu\nu}=-\frac{1}{2}\Box_{d+1}\delta g_{\mu\nu}+R_{\rho\mu\sigma\nu}\delta g^{\rho\sigma}-\frac{1}{2}R_{\mu\rho}\delta g^\rho_\nu-\frac{1}{2}R_{\nu\rho}\delta g^\rho_\mu\equiv \Delta_L\delta g_{\mu\nu}
\end{equation}
where $\Box_{d+1}$ is the ($d+1$)-dimensional covariant d'Alembertian, $R_{\mu\nu\rho\sigma}$ is the Riemann tensor and $\Delta_L$ is the Lichnerowicz operator. 

Introducing the adimensional coordinates discussed in Section \ref{gravloc} (that we will use in the rest of this section), using the explicit expression for the Christoffels and the components of the Riemann and Ricci tensors and defining the covariant Laplacian $\Delta_{d-1}$ on the unit ($d-1$)-sphere, after a long calculation the spatial components (all the other components are identically satisfied) of equation (\ref{pert}) can be recast in the form
\begin{equation}
\begin{split}
-\frac{1}{f(y)}\partial_{\tilde{t}}^2\delta g_{kl}+f(y)\partial_y^2\delta g_{kl}+\left[f'(y)-\frac{(5-d)f(y)}{y}\right]\partial_y\delta g_{kl}+\frac{1}{y^2}\Delta_{d-1}\delta g_{kl}\\
+2\left[\frac{2f(y)}{y^2}-\frac{d-1}{y^2}\right]\delta g_{kl}
=\left[\frac{2d}{\gamma^2}-\frac{2(d-2)q^2}{y^{2d-2}}\right]\delta g_{kl}.
\end{split}
\label{waveeq2}
\end{equation}
We can now decompose the metric perturbation in terms of the tensor harmonics on the unit ($d-1$)-sphere (defined in Section \ref{gravloc}):
\begin{equation}
\delta g_{kl}=\sum_k y^{\frac{5-d}{2}}\phi_k(\tilde{t},y)\mathbb{T}_{kl}^{(k)}
\label{decomp}
\end{equation}
and introduce the tortoise coordinate $dr^*=dy/f(y)$, in terms of which the horizon of the black hole is mapped to $r^*\to -\infty$ and the asymptotic boundary to a finite value $r^*=r^*_\infty$. Equation (\ref{waveeq2}) reduces then to a wave equation for each of the modes $\phi_k(\tilde{t},y)$:
\begin{equation}
-\partial_{\tilde{t}}^2\phi_k(\tilde{t},y)=-\partial_{r^*}^2\phi_k(\tilde{t},y)+V_k\left[y(r^*)\right]\phi_k(\tilde{t},y)
\label{wave}
\end{equation} 
where the potential $V_k\left[y(r^*)\right]$ is given by equation (\ref{potential}) and is in accordance with the results of ref\cite{kodama}. In the frequency domain, equation (\ref{wave}) finally takes the form of the Schr\"odinger equation (\ref{schro}), which encodes the effects of the radial extra dimension on a transverse-traceless graviton.

By finding the expression of the tortoise coordinate $r^*$ as a function of the adimensional radial coordinate $y$ in the limit of large $y$, it is easy to show that the potential (10) diverges at the asymptotic boundary as $V(r^*)\sim 15/[4(r^*_\infty-r^*)^2]$. This behaviour is very similar to the one typical of the RS II potential\cite{RS1,RS2,karchrandall}. On the other hand, at the black hole horizon ($r^*\to -\infty$) the potential vanishes exponentially: $V(r^*)\sim \exp(2k_gr^*)$ where $k_g$ is the above defined surface gravity of the black hole. Since the normalizable zero-graviton mode bound on the brane present in the RS II model is a marginally bound mode, and since our potential vanishes exponentially while RS's one exhibited a power-law behaviour deep into the bulk, we cannot expect to find a normalizable bound mode as well. But the presence of the brane still allows the existence of a quasi-bound mode, with complex frequency $\omega=\bar{\omega}+i\Gamma/2$, very well localized close to the brane. Its finite lifetime (determined by the imaginary part of the frequency $\Gamma/2$) implies that it will stay on the brane for an interval of time, after which it will eventually leak into the bulk black hole.

Up to this point, we only studied the perturbed Einstein equations, without taking into account the presence of the ETW brane. To study the evolution of a TT graviton living on the brane, and to understand whether or not it is possible for this graviton to remain bound on the brane at least locally and for a reasonable amount of time, we need to enforce a boundary condition for the graviton wavefunction at the position of the brane. At each instant of time, the brane is effectively a hypersurface at fixed $y$ (i.e. at fixed $r^*$), therefore the previous analysis can be applied to study a gravitational perturbation on the brane. In particular, the Schr\"odinger equation allows us to understand if the graviton is bound on the brane or if it necessarily leaks into the bulk. If the brane is expanding or contracting, it is not clear how to implement such a boundary condition. But if the time scale of the perturbation is much smaller than the time scale of the motion of the brane, the linearization of equation (\ref{brane}) provides the boundary condition needed. In other words, we must work under an adiabatic approximation for the motion of the brane with respect to the time scale of the perturbation, so that we can approximately consider the ETW brane to be static at $y=y_b$, or $r^*=r^*_b$. Additionally, since all the calculations will be carried out in the bulk coordinates assuming the brane to be static, in order for an observer comoving with the brane to interpret correctly the graviton, we must also verify that the redshift between the bulk coordinates and the comoving coordinates on the brane is compatible with the assumption of a static brane. From the definition of proper time, this means that $y'^2\ll f^2(y)$ must hold.

\begin{figure}[H]
\centering
\includegraphics[height=4cm]{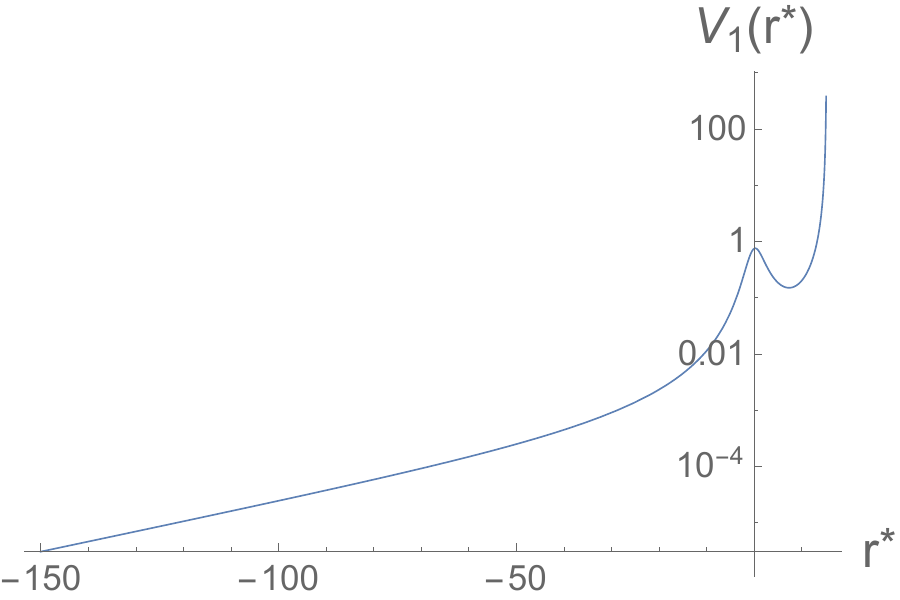}
\caption{\textbf{Potential $\mathbf{V_k[y(r^*)]}$ - Small black hole.} $r_+=0.1$, $r_-=0.099$, $L_{AdS}=1$, $l=k=1$. The potential diverges for $r^*_\infty=15.3$ and vanishes exponentially at the horizon $r^*\to -\infty$. Differently from the large black hole case (see Figure \ref{fig4}), the potential exhibits a peak for $r^*\sim 0$.}
\label{potsmall}
\end{figure}

Equation (\ref{brane}) can be traced and rewritten as
\begin{equation}
K_{ab}=\tilde{T}h_{ab}
\label{extrbound}
\end{equation}
where $\tilde{T}=r_+T$ and the extrinsic curvature is computed using the adimensional coordinates. For a static brane, the normal dual vector is given by $n_\mu=\delta_{\mu,y}/\sqrt{f(y)}$ and $e^\mu_a=\delta^\mu_a$. Linearizing equation (\ref{extrbound}) and using the decomposition (\ref{decomp}) of the metric perturbation, we obtain the (Neumann) boundary condition at the brane
\begin{equation}
\partial_y\psi_{k,\omega}\Big|_{y=y_b}=\left.\frac{4\tilde{T}y-(5-d)\sqrt{f(y)}}{2y\sqrt{f(y)}}\psi_{k,\omega}\right|_{y=y_b}.
\end{equation}
Using the condition $y'^2\ll f^2(y)$, equation (\ref{traj}) guarantees that $f(y)\sim \tilde{T}^2y^2$. The latter result and the definition of the tortoise coordinate finally lead to the boundary condition reported in Section \ref{gravloc}, i.e.
\begin{equation}
\partial_{r^*}\psi_{k,\omega}\Big|_{r^*=r^*_b}=\left.\frac{(d-1)\sqrt{f\left[y(r^*)\right]}}{2y(r^*)}\psi_{k,\omega}\right|_{r^*=r^*_b}.
\label{finalbound}
\end{equation}
As we have already discussed, imposing this boundary condition is equivalent to add to the potential (10) a negative delta term, which ``traps'' the graviton, allowing gravity localization. Before reviewing the trapping coefficient method introduced in ref.\cite{resonances} and give more details about our numerical results, we remark that in the small black hole case ($\gamma \gtrsim 1$) the potential (10) presents a peak for $r^*\sim 0$ before diverging at $r^*=r^*_\infty$ (see Figure \ref{potsmall}). This feature disappears for large black holes. If the brane radius was able to reach values consistently larger than the position of the peak, it would have implied not only the presence of a quasi-bound zero mode, but also the existence of ``overtones''\cite{resonances} trapped in the resonance cavity between the peak and the brane position. Nonetheless, as we have already shown, in the small black hole case the ratio between the maximum brane radius and the horizon of the black hole cannot be large, implying that no quasi-bound modes exist. Therefore, the small black hole case is not of our interest and we will focus our attention on the $\gamma\ll 1$ case.

\subsection{Trapping coefficient method}

The quasi-normal modes, already studied in ref.\cite{resonances} for the Schwarzschild-AdS spacetime in the presence of a static brane, are modes with a finite lifetime determined by the imaginary part of the frequency $\Gamma/2$ and purely infalling boundary condition at the black hole horizon, i.e. $\psi_{k,\omega}\sim \exp(i\omega r^*)$ for $r^*\to -\infty$. This finite lifetime can be understood as a probability of tunneling through the potential barrier from the delta potential at the position of the brane toward the black hole horizon. The absence of the resonant cavity in the potential (10) for the large black hole case implies that only one resonant mode, supported by the attractive delta potential, exists, that we will call quasi-bound mode. Our analysis is based on the trapping coefficient method introduced in ref.\cite{resonances}, that we will review and apply to the AdS-Reissner-Nordstr\"om case.

Since the potential vanishes approaching the horizon, the wavefunction in that region must take the form of a plane wave:
\begin{equation}
\psi_{k,\omega}(r^*)\sim \frac{1}{2}A_h\textrm{e}^{-i\delta(\omega)}\left[\textrm{e}^{-i\omega r^*}+S(\omega)\textrm{e}^{i\omega r^*}\right]
\label{horsol}
\end{equation}
where $S(\omega)=\textrm{e}^{2i\delta(\omega)}$ is the scattering matrix and $\delta(\omega)$ is the scattering phase shift. The black hole horizon is located at $r^*=-\infty$, and we want purely infalling solutions, therefore in that limit $\psi_{k,\omega}$ must be a left-moving plane wave. Since we are imposing two boundary conditions (equation (\ref{finalbound}) at the brane and the pure infalling one at the horizon), we expect to find a discrete set of solutions to the Schr\"odinger equation (the quasi-normal modes), corresponding to a discrete set of frequencies $\omega_n$. By considering complex frequencies, we can obtain the purely infalling solutions by requiring the scattering matrix to have a pole at the quasi-normal mode frequencies $\omega_n$\footnote{In the large black hole case, only one mode, supported by the attractive delta potential, will be present in the spectrum. In the following we will then drop the $n$ index, understood to take the value $n=0$.}. It can be shown \cite{resonances,scattering} that the leading order Laurent expansion of $S(\omega)$ is given by:
\begin{equation}
S(\omega)\sim \textrm{e}^{2i\delta_0(\omega)}\frac{\omega-\omega_n^*}{\omega-\omega_n}
\end{equation}
where $\delta_0(\omega)$ is a slowly varying real function of $\omega$. Using the definition $\omega_n=\bar{\omega}_n+i\Gamma_n/2$, the scattering phase shift can be expressed as:
\begin{equation}
\delta(\omega)\sim \delta_0(\omega)+\arcsin\left[\frac{\Gamma_n}{\sqrt{4(\omega-\bar{\omega}_n)^2+\Gamma_n^2}}\right].
\label{flipping}
\end{equation}
For real $\omega$ and if $\Gamma_n$ is small with respect to $\bar{\omega}_n$, $\delta(\omega)$ varies of a value which is approximately $\pi$ when $\omega$ is varied across the real part of the resonance frequency, flipping the sign of the wavefunction (\ref{horsol}). We can now define the trapping coefficient
\begin{equation}
\eta(\omega)\equiv\frac{A_b}{A_h}
\label{trapping}
\end{equation}
where $A_b$ is the magnitude of the wavefunction $\psi_{k,\omega}$ at the brane and $A_h$ its magnitude at the horizon of the black hole (effectively, in a region where the potential is almost vanishing). Intuitively, since we expect the quasi-bound mode to be almost localized on the brane due to the presence of the attractive delta potential at $r^*=r_b^*$, the trapping coefficient will present a peak at the frequency of the quasi-bound mode. In the case of a normalizable bound mode (with vanishing imaginary part of the frequency), clearly the wavefunction would vanish at the horizon, giving an infinite trapping coefficient. For a purely infalling solution we can write in general\cite{resonances}:
\begin{equation}
\psi_{k,\omega}(r^*)=N(\omega)\begin{cases}
A_h\Re\left[\textrm{e}^{i\delta(\omega)}\textrm{e}^{i\omega r^*}\right] \hspace{0.5cm} r^*\to-\infty\\[10pt]
R(\omega)\Re\left[\textrm{e}^{i\delta(\omega)}\textrm{e}^{i\theta(\omega)}\right]=-R(\omega)\sin\left(\delta(\omega)+\theta(\omega)-\frac{\pi}{2}\right)\equiv A_b\hspace{0.5cm} r^*=r^*_b
\end{cases}
\end{equation}
where $\Re$ indicates the real part and $R(\omega)$ and $\theta(\omega)$ are slowly-varying functions of the frequency. For every frequency $\omega$ we can choose the normalization $N(\omega)$ of the wavenfunction such that $A_b=1$, and of course the trapping coefficient will be unchanged. This allows us to set $\psi(r^*_b)=1$ in our numerical analysis, obtaining:
\begin{equation}
\psi_{k,\omega}(r^*)=\begin{cases}
-\frac{A_h}{R(\omega)\sin\left(\delta(\omega)+\theta(\omega)-\frac{\pi}{2}\right)}\Re\left[\textrm{e}^{i\delta(\omega)}\textrm{e}^{i\omega r^*}\right]=-\frac{1}{\eta(\omega)}\Re\left[\textrm{e}^{i\delta(\omega)}\textrm{e}^{i\omega r^*}\right]\hspace{0.5cm} r^*\to-\infty\\[10pt]
1 \hspace{0.5cm} r^*=r^*_b
\end{cases}
\end{equation}
where the trapping coefficient therefore reads:
\begin{equation}
\eta(\omega)=\frac{R(\omega)\sin\left(\delta(\omega)+\theta(\omega)-\frac{\pi}{2}\right)}{A_h}.
\end{equation}
If the relation
\begin{equation}
\delta_0(\omega)+\theta(\omega)\sim\frac{\pi}{2},\frac{3\pi}{2}
\label{condlorentzian}
\end{equation}
holds, the square of the trapping coefficient takes the form:
\begin{equation}
\xi(\omega)\equiv\eta^2(\omega)\sim\frac{R^2(\omega)}{A_h^2}\frac{\Gamma_n^2}{4(\omega-\bar{\omega}_n)^2+\Gamma_n^2}.
\label{lorentziancurve}
\end{equation}
This Lorentzian (Breit-Wigner) peak is centred at the real part of the frequency of the quasi-bound mode, while its half-width at half-maximum $\Gamma/2$ gives the imaginary part of the frequency, and therefore the lifetime of the mode. If the condition (\ref{condlorentzian}) is not satisfied, the shape of the peak is more complicated. Nonetheless, at least for the zero mode of our interest, it has been shown that the condition (\ref{condlorentzian}) holds in a good approximation\cite{resonances} and our numerical analysis confirms this result. We mention here the possibility of refining such an approximation by subtracting from the plot of $\xi(\omega)$ as a function of the frequency a baseline function accounting for the slow variation of $\delta_0(\omega)$ and $\theta(\omega)$ with the frequency. For our purposes, such a procedure (which leads to some difficulties due to the arbitrariness of the choice of the baseline\cite{resonances}), turns out to be not necessary. It is therefore possible, in a reasonable approximation, to find the real and imaginary parts of the frequencies of the quasi-bound mode (we remind that for the large black hole case we expect only the $n=0$ mode to be present, supported by the negative delta potential at the position of the brane) using the following procedure (trapping coefficient method):
\begin{enumerate}
\item Compute numerically a family of solutions of the Schr\"odinger equation (\ref{schro}), parametrized by real-valued frequencies $\omega$ and with boundary condition (\ref{finalbound}), requiring also $\psi_{k,\omega}(r_b^*)=1$;
\item Find numerically the maximum of the wavefunction in the region where the potential is almost vanishing (i.e. near the horizon) for a range of values of $\omega$;
\item Plot the square of the inverse of these maxima as a function of $\omega$: a peak will be present at the real part of the frequency of the (purely infalling) quasi-normal mode;
\item Eventually subtract a baseline function;
\item Fit the data with the Breit-Wigner distribution (\ref{lorentziancurve}) to find real and imaginary parts of the frequency of the quasi bound mode.
\end{enumerate}

\subsection{Numerical results}

We will find the real and imaginary parts of the frequency and the height of the peak $R/A_h$ for three sizes of the black hole: $r_+=10,100,10000$ (with $L_{AdS}=1$ and $r_-$ such that the black holes are near extremality), corresponding to $\gamma=0.1,0.01,0.0001$ respectively. For each case we will consider different positions $y_b$ for the brane and values for the angular momentum $l$. The results of the fits are reported in Table \ref{fitresults}. 
\begin{table}[H]
\centering\scalebox{1}{
\begin{tabular}{|c|c|c|c|c|c|c|c|}
$\gamma$ &$y_b$ &$l$ &$r_-$ &$R/A_h$ &$\bar{\omega}$ &$\Gamma$ &$\omega_{GR}$\\ \hline
0.1 &17.34 &1 &9.99 &$68.22$ &17.22 &0.7354 &17.32\\ \hline
0.1 &17.34 &5 &9.99 &$61.41$ &58.93 &0.9186 &59.16\\ \hhline{|=|=|=|=|=|=|=|=|}
0.01 &17.34 &1 &99.9 &$71.51$ &168.2 &67.22 &173.2\\ \hline
0.01 &34.62 &1 &99.9 &$202.5$ &147.5 &16.86 &173.2\\ \hline
0.01 &60.22 &1 &99.9 &$464.8$ &71.70 &5.580 &173.2\\ \hline
0.01 &63 &1 &99.9 &$499.2$ &52.40 &5.060 &173.2\\ \hline
0.01 &66 &1 &99.9 &528.8 &12.56 &4.835 &173.2\\ \hline
0.01 &66.15 &1 &99.9 &$\sim 519.6$ &$\sim 4$ &$\sim 5$ &173.2\\ \hline
0.01 &17.34 &5 &99.9 &70.74 &586.4 &68.55 &591.6\\ \hline
0.01 &30.00 &5 &99.9 &160.9 &585.0 &23.09 &591.6\\ \hline
0.01 &34.62 &5 &99.9 &$199.5$ &583.6 &17.28 &591.6\\ \hline
0.01 &99.94 &5 &99.9 &$980.9$ &530.5 &2.081 &591.6\\ \hline
0.01 &60.00 &10 &99.9 &447.8 &$1.084\cdot 10^3$ &5.984 &$1.095\cdot 10^3$\\ \hline
0.01 &120.0 &20 &99.9 &$1.229\cdot 10^3$ &$2.074\cdot 10^3$ &1.587 &$2.098\cdot 10^3$\\
\hhline{|=|=|=|=|=|=|=|=|}
0.001 &17.34 &5 &999 &$\sim 73$ &$\sim 4\cdot 10^3$ &$\sim 5\cdot 10^3$ &$5.916\cdot 10^3$\\
\hhline{|=|=|=|=|=|=|=|=|}
0.0001 &34.82 &100 &9900 &$203.2$ &$4.181\cdot 10^5$ &$1.737\cdot 10^5$ &$1.010\cdot 10^6$\\ \hline
0.0001 &34.82 &1000 &9900 &$180.9$ &$9.955\cdot 10^6$ &$2.097\cdot 10^5$ &$1.001\cdot 10^7$\\ \hline
0.0001 &34.82 &10000 &9900 &$160.9$ &$9.993\cdot 10^7$ &$2.552\cdot 10^5$ &$1.000\cdot 10^8$\\ \hline
\end{tabular}}
\caption{\textbf{Quasi-bound modes.} Results of the numerical analysis based on the trapping coefficient method. We set the AdS radius to be $L_{AdS}=1$. In the first column we report the size of the black hole $\gamma=L_{AdS}/r_+$, in the second column the position of the brane in adimensional coordinates $y_b=r_b/r_+$, in the third column the angular momentum $l$, in the fourth column the value of the inner horizon radius $r_-$. In the fifth, sixth and seventh column we show the results of the Breit-Wigner fits of the squared trapping coefficient: the amplitude $R/A_h$ of the trapping coefficient and the real ($\bar{\omega}$) and imaginary ($\Gamma/2$) parts of the frequency of the quasi-bound mode, respectively. In the last column we show the expected frequency $\omega_{GR}$ of a 4-dimensional perturbation of an Einstein static universe. The fitted values for $\gamma=0.01$, $y_b=66.15$, $l=1$ and for $\gamma=0.001$, $y_b=17.34$, $l=5$ are purely indicative. We remark that for each set of parameters reported above it is possible to choose a tension $T$ for the brane such that the corresponding $y_b$ is, for instance, the turning point of the brane, while the corresponding Euclidean solutions are sensible ($\tau_0>0$) and dominant in the gravity path integral ($\Delta I<0$).}
\label{fitresults}
\end{table}
\noindent From our analysis the following picture emerges (we remind that $\gamma=L_{AdS}/r_+$ and $y_b=r_b/r_+$):
\begin{itemize}
\item For fixed $\gamma$ and $l$, increasing the distance of the brane $y_b$ the real part of the frequency $\bar{\omega}$ decreases and the peak of the square of the trapping coefficient $\xi$ becomes sharper and higher, indicating that the imaginary part of the frequency $\Gamma/2$ decreases as well and gravity localization is more efficient. This is evident from the data reported in Table \ref{timescales} (where $t_o=1/\bar{\omega}$ and $t_d=2/\Gamma$). But if the brane is very far, $\bar{\omega}$ approaches zero. When it happens, the peak becomes broader (indicating $\Gamma\sim \bar{\omega}$) and is not approximable with a Lorentzian curve anymore. At this point, the peak is very high, indicating a short-lived (large ratio between imaginary and real part of the frequency) but very well localized quasi-bound state. After that, for even farther brane, the quasi-bound mode is no more present (there is no peak, and the trapping coefficient never becomes very big). Additionally, for low values of the angular momentum $l$, the real part of the frequency obtained is not in accordance with the one expected from a 4-dimensional perturbation of an Einstein static universe (given by $\omega_{GR}=\sqrt{f(y_b)\cdot(l+2)l}/y_b$)\cite{resonances}, even when the brane is far and the peak is narrow. For higher values of $l$ instead, i.e. for smaller scale perturbations, the GR expected value and the one obtained numerically are closer. We report here the data showing these behaviours only in the $\gamma=0.01$ case, but we verified that the same analysis applies also for $\gamma=0.1,0.0001$ as well as in the small black hole case. The smaller is the black hole, the farther is the position of the brane where gravity localization is lost. We remark how this behaviour is in contradiction with the conclusions of ref.\cite{resonances}, where it is argued that the real part of the frequency approaches a constant value and the RS II normalizable bound zero-graviton mode is recovered in the far brane limit. However, it is worth noting that the loss of localization for the small black hole case, which is the one mainly studied in ref.\cite{resonances}, occurs for a position of the brane much farther than the ones explored in Figure 11 of ref.\cite{resonances}
\begin{figure}[H]
\centering
\resizebox{\textwidth}{!}{
    \begin{tabular}{ccc}
\includegraphics[height=4cm]{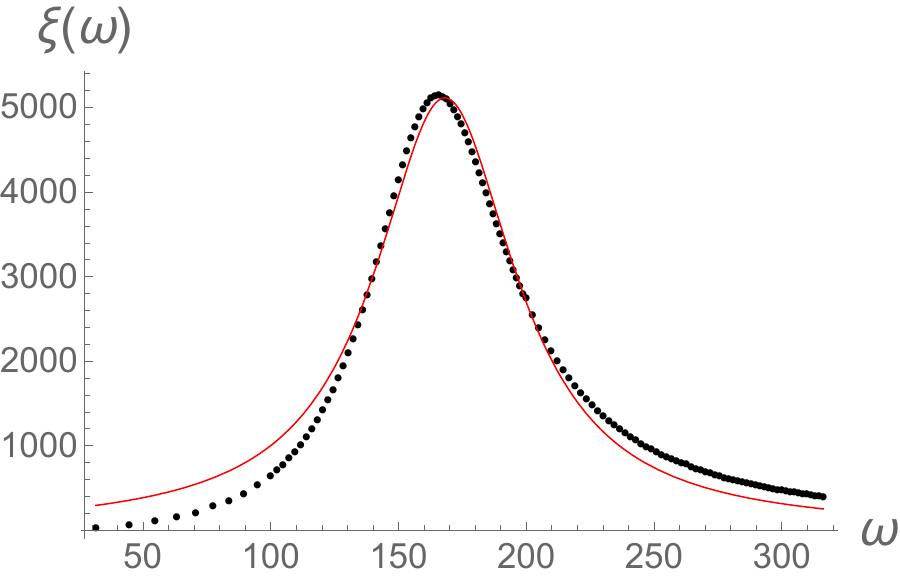}

    &   \includegraphics[height=4cm]{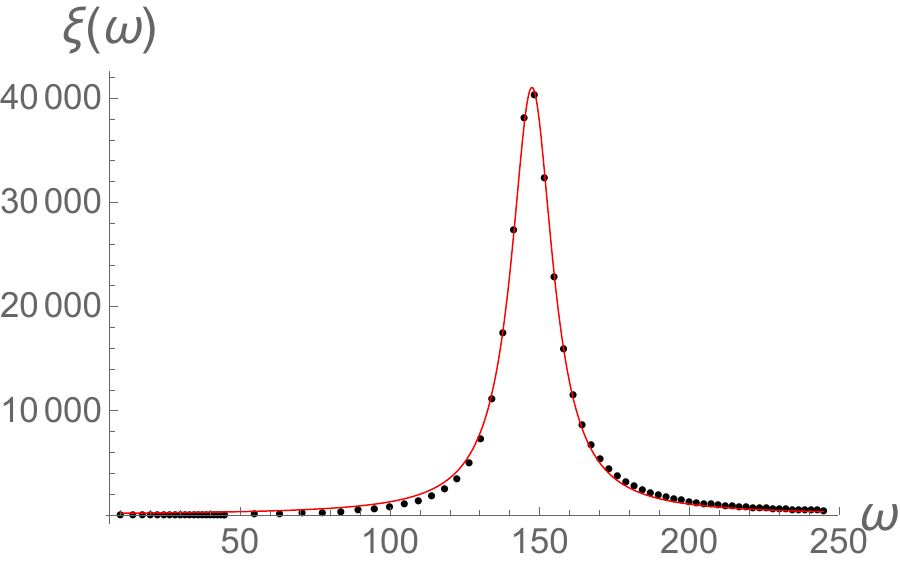}
         &   \includegraphics[height=4cm]{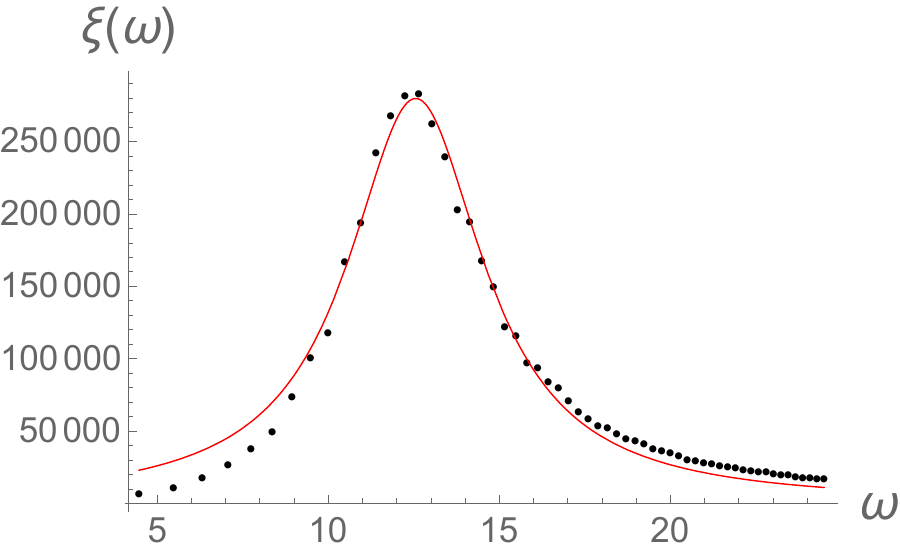}\\
         (a) & (b) &(c)
   \end{tabular}
}
\caption{\textbf{Delocalization for far brane.} $d=4$, $r_+=100$ ($\gamma=0.01$), $r_-=99.9$, $L_{AdS}=1$, $l=1$. For fixed black hole size and angular momentum, if the brane is too far from the horizon gravity localization is lost. (a) $y_b=17.34$. If the brane is not far enough from the horizon, the peak of the squared trapping coefficient is not well approximated by the Lorentzian profile and gravity localization is not very efficient. (b) $y_b=34.62$. When the brane is farther, gravity localization is more efficient and the peak of the squared trapping coefficient is Lorentzian. (c) $y_b=66$. If the brane is too far from the horizon, the peak becomes broad, the mode is short-lived and gravity is no more localized.}
\end{figure}
\item For fixed $l$ and $y_b$, decreasing $\gamma$ (i.e. increasing the size of the black hole) the real part of the frequency grows and the peak becomes broader, indicating a shorter-lived quasi-bound mode. The Lorentzian approximation of the peak is also less precise. At the same time, we remind that the same maximum radius of the brane $y_b^{max}=r_0/r_+$ is reached for a smaller value of the tension $T$ if the black hole is larger.
\begin{figure}[H]
\centering
\resizebox{\textwidth}{!}{
    \begin{tabular}{ccc}
\includegraphics[height=4cm]{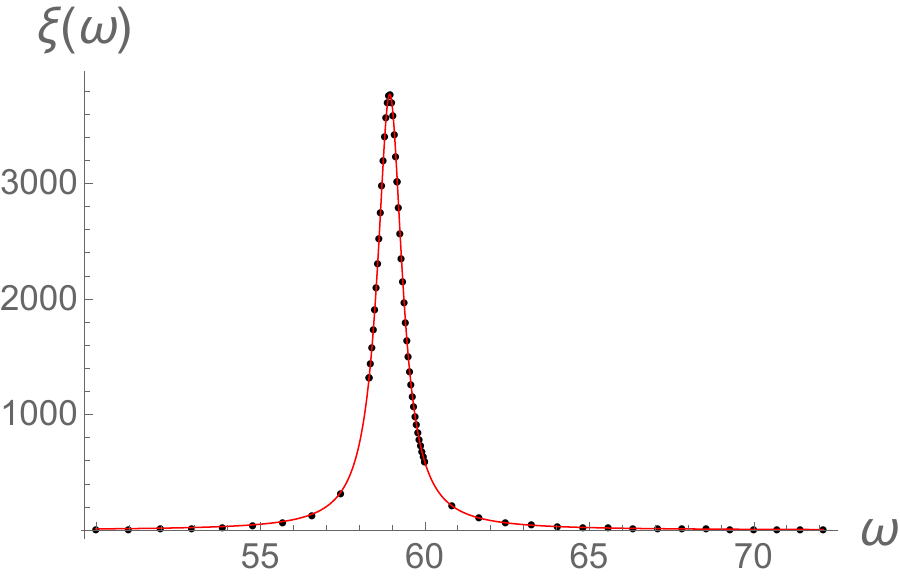}

    &   \includegraphics[height=4cm]{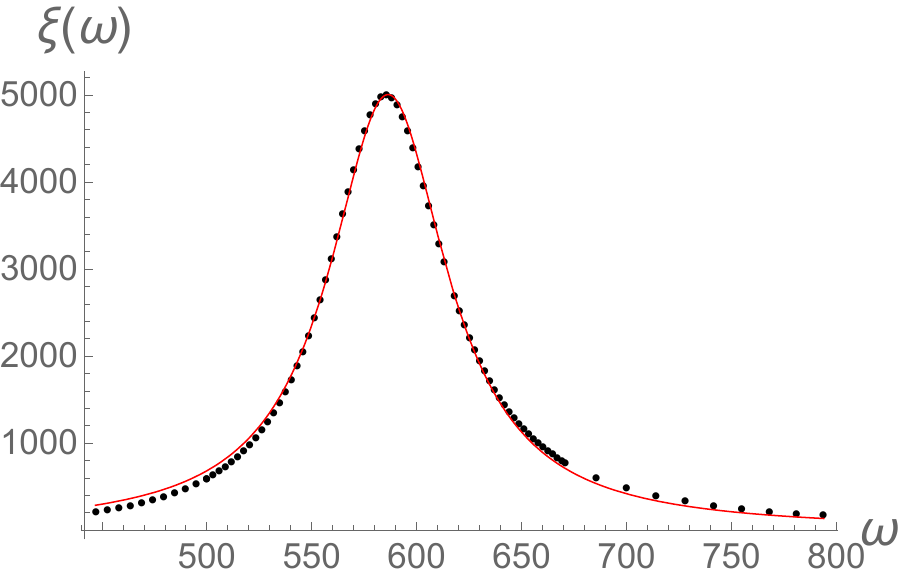}
    &  \includegraphics[height=4cm]{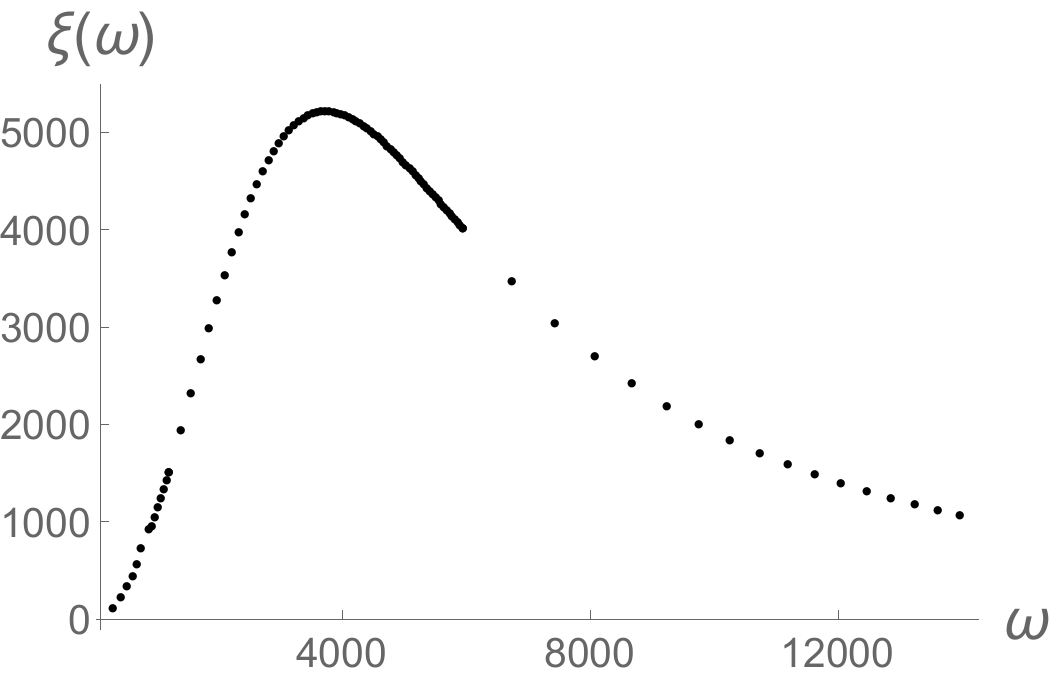}\\
         (a) & (b) &(c)
   \end{tabular}
}
\caption{\textbf{Delocalization for large black holes.} $d=4$, $r_-/r_+=0.99$, $L_{AdS}=1$, $y_b=17.34$, $l=5$, (a) $\gamma=0.1$, (b) $\gamma=0.01$ and (c) $\gamma=0.001$. Increasing the size of the black hole for fixed angular momentum and radius of the brane, the peak of the squared trapping coefficient is broader and not well approximated by the Lorentzian profile: gravity localization becomes less efficient.}
\end{figure}
\item Holding $\gamma$ and $y_b$ fixed and increasing the angular momentum $l$ the real part of the frequency grows. In the small black hole case, we verified that the imaginary part of the frequency decreases, exactly in the same way as it was described in ref.\cite{resonances} Differently, in the large black hole case, the imaginary part grows, but much slower than the real part. The peaks are therefore narrower and the ratio between the imaginary and the real part decreases. At the same time, since the imaginary part of the frequency increases, the height of the peaks decreases with increasing $l$, indicating that, even if they are able to undergo a larger number of oscillations before decaying, these smaller-scale modes have a shorter lifetime than the larger-scale ones (even if localization is still very efficient). These characteristics can be observed comparing the peaks for different values of $l$ in all the three cases, but are particularly evident in the very large black hole cases $\gamma=0.0001$. We remark that even for a very far brane, which does not support localization of gravity for instance for $l=1$, increasing the angular momentum the quasi-bound state is recovered. For example, for $\gamma=0.01$ and $l=1$ the quasi bound state is lost for $y_b\sim 66$, but for $l=5$ it is still present when the brane radius is $y_b=99.94$.
\begin{figure}[H]
\centering
\resizebox{\textwidth}{!}{
    \begin{tabular}{cc}
\includegraphics[height=4cm]{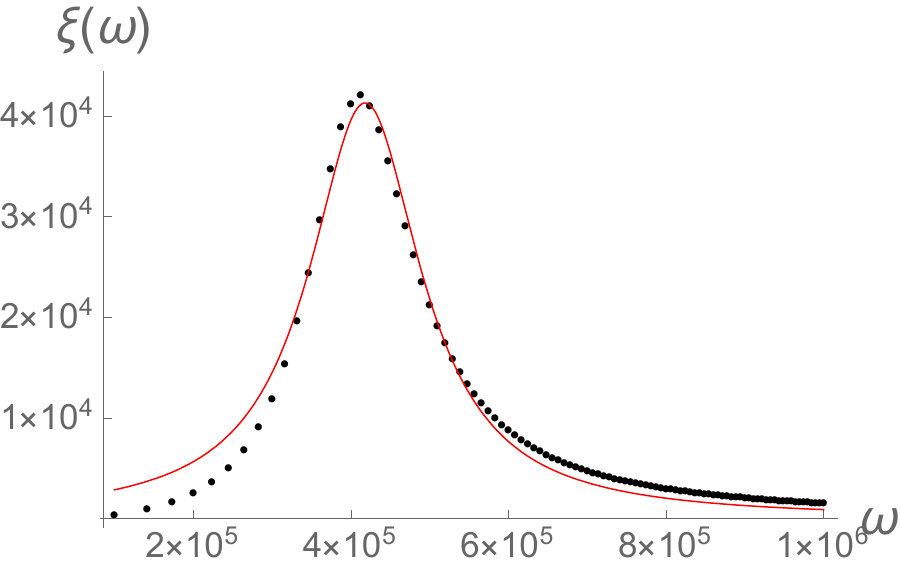}

    &   \includegraphics[height=4cm]{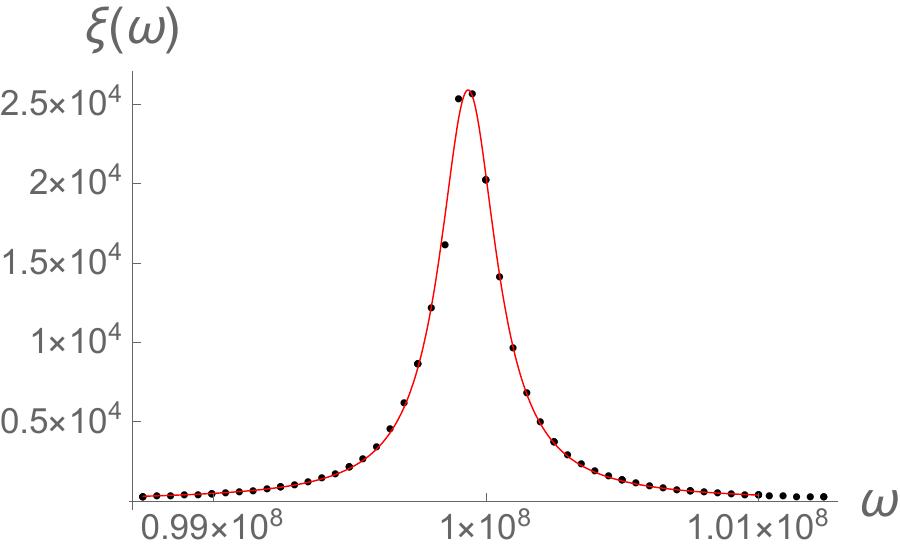}\\
         (a) & (b)
   \end{tabular}
}
\caption{\textbf{Localization for small scales.} $d=4$, $L_{AdS}=1$, $r_+=10000$ ($\gamma=0.0001$), $r_-=9900$, $y_b=34.62$, (a) $l=100$ and (b) $l=10000$. For fixed radius of the brane and size of the black hole, increasing the angular momentum (i.e. decreasing the scale of the perturbation) the peak of the squared trapping coefficient is better approximated by the Lorentzian profile and gravity localization becomes more efficient.}
\end{figure}
\item Keeping the size of the black hole and the scale of the perturbation constant while increasing the brane radius, i.e. increasing $l$ and $y_b$ of the same factor holding $\gamma$ fixed, the peak becomes narrower and gravity localization more efficient: remaining on a local enough scale, if the brane is farther we are in a setup more similar to the RS II scenario.
\begin{figure}[H]
\centering
\resizebox{\textwidth}{!}{
    \begin{tabular}{ccc}
\includegraphics[height=4cm]{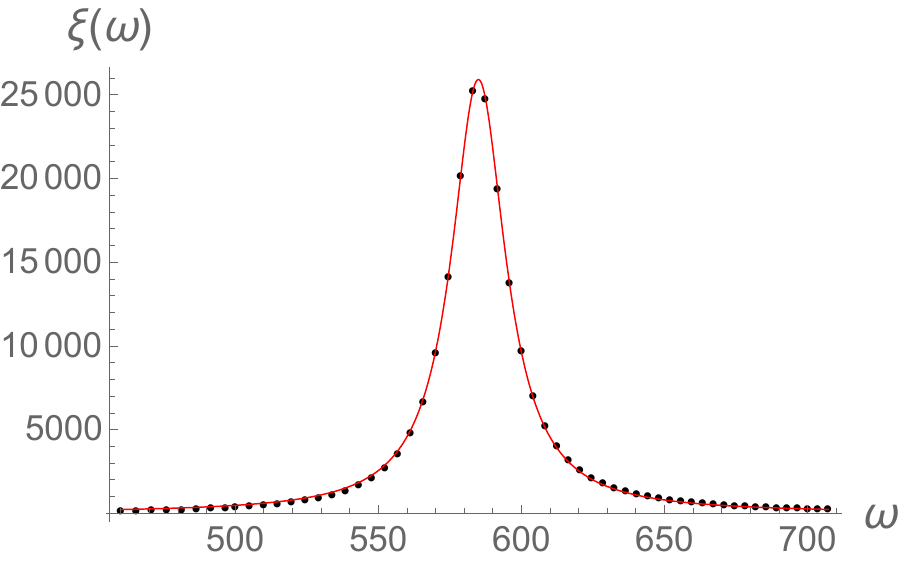}

    &   \includegraphics[height=4cm]{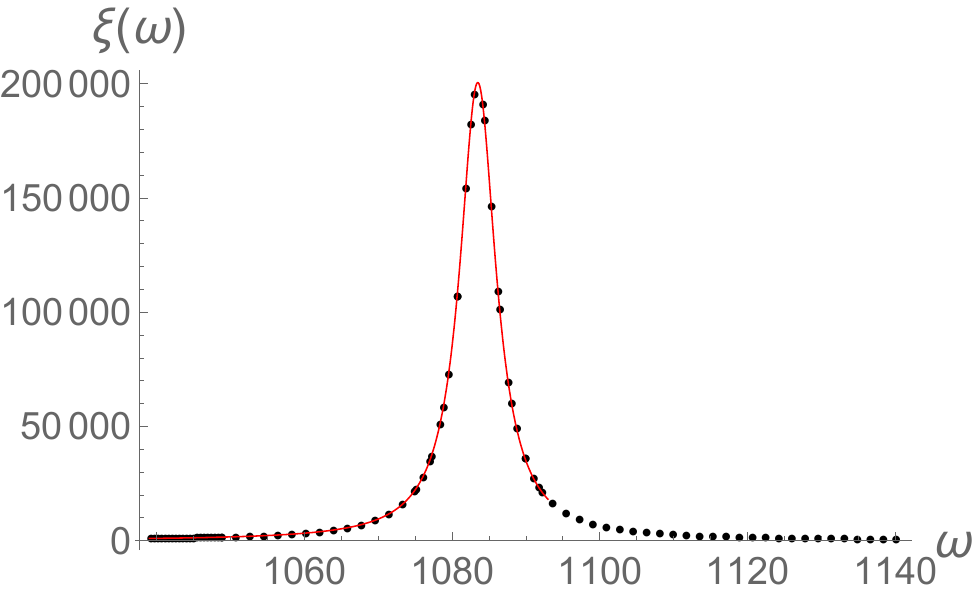}
         &   \includegraphics[height=4cm]{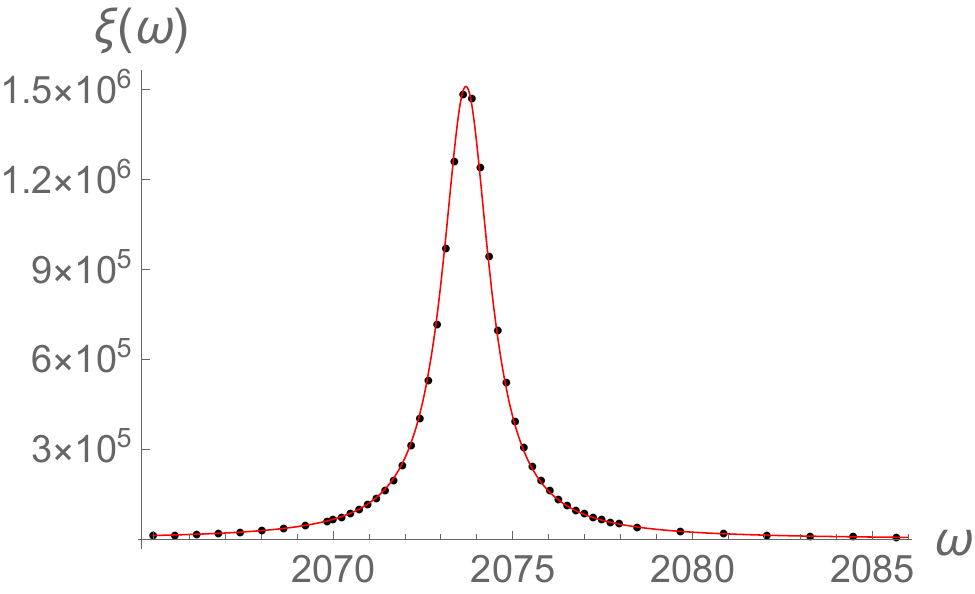}\\
         (a) & (b) &(c)
   \end{tabular}
}
\caption{\textbf{Localization for far brane and small scales.} $d=4$, $L_{AdS}=1$, $r_+=100$ ($\gamma=0.01$), $r_-=99.9$. (a) $l=5$, $y_b=30$. (b) $l=10$, $y_b=60$. (c) $l=20$, $y_b=120$. For fixed black hole size, increasing the angular momentum and the radius of the brane of the same factor, the mode is longer-lived and gravity localization becomes more efficient.}
\end{figure}
\end{itemize}
\begin{figure}[H]
\centering
\resizebox{\textwidth}{!}{
    \begin{tabular}{ccc}
\includegraphics[height=4cm]{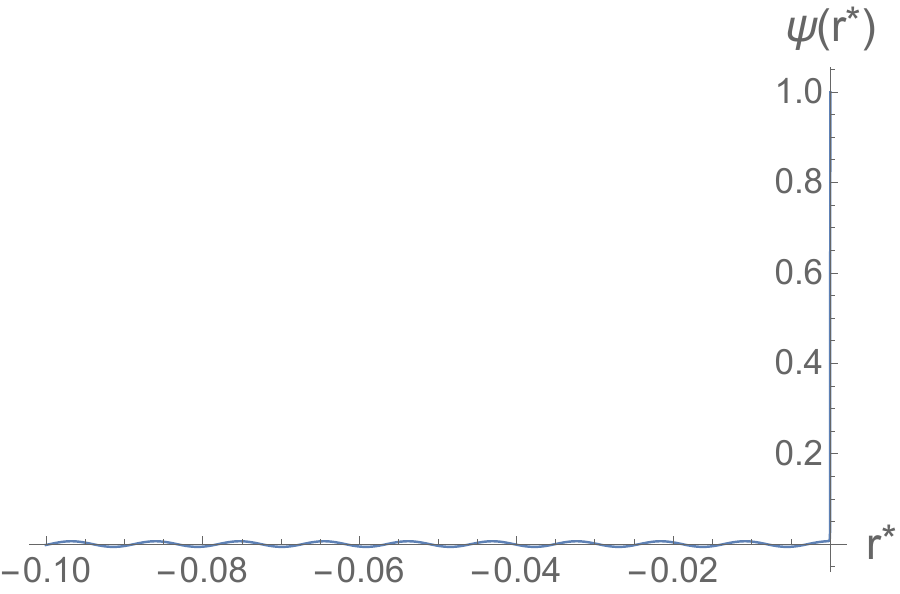}

    &   \includegraphics[height=4cm]{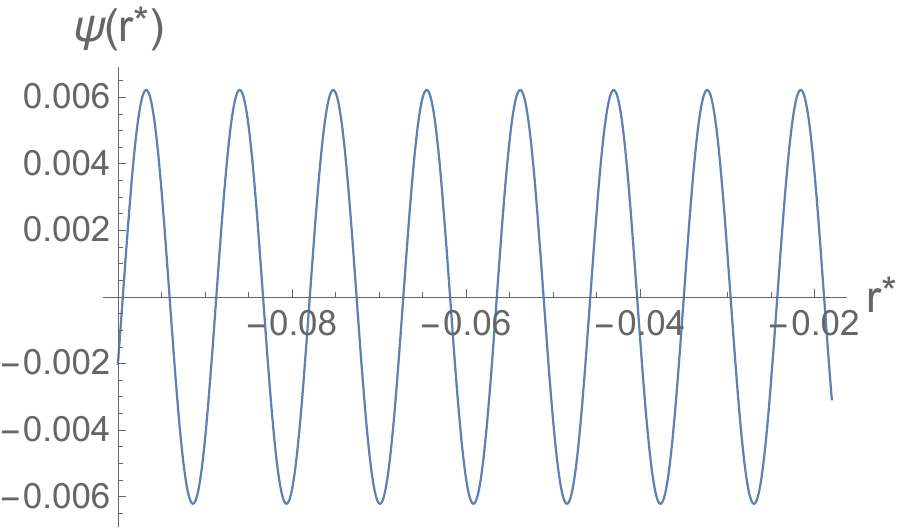}
         &   \includegraphics[height=4cm]{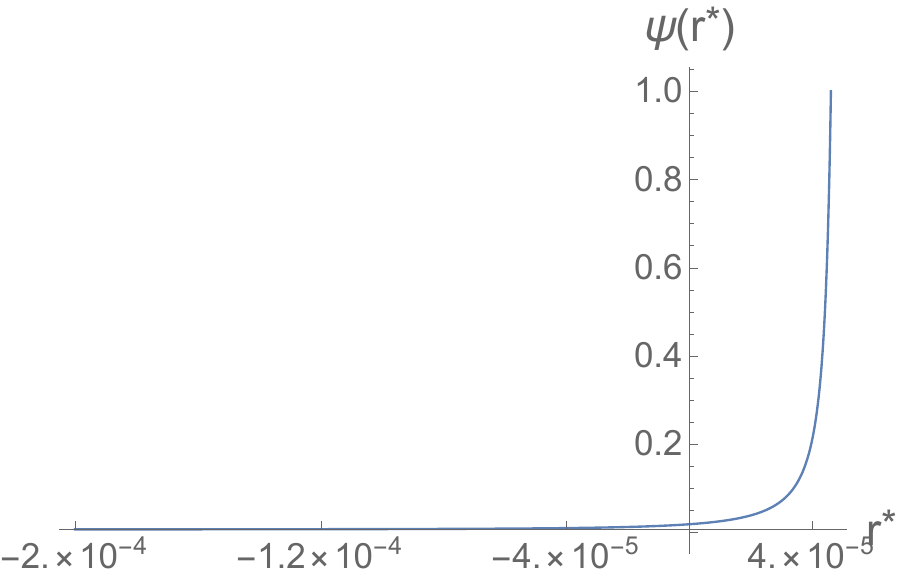}\\
         (a) & (b) &(c)
   \end{tabular}
}
\caption{\textbf{Example of quasi-bound mode wavefunction.} $d=4$, $L_{AdS}=1$, $r_+=100$ ($\gamma=0.01$), $r_-=99.9$, $l=5$, $y_b=30$. (a) Quasi-bound mode wavefunction. (b) Detail of the oscillatory behaviour near the horizon. (c) Detail of the growth near the brane.}
\end{figure}

\subsection{Time scales and adiabatic approximation}

\begin{figure}[H]
\centering
\resizebox{\textwidth}{!}{
    \begin{tabular}{ccc}
\includegraphics[height=4cm]{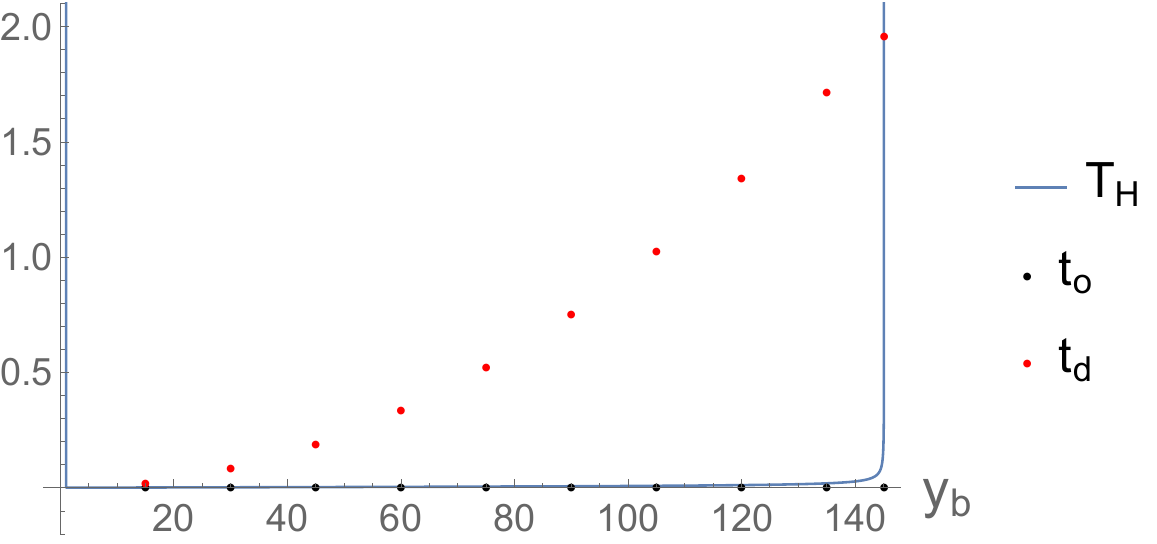}

    &   \includegraphics[height=4cm]{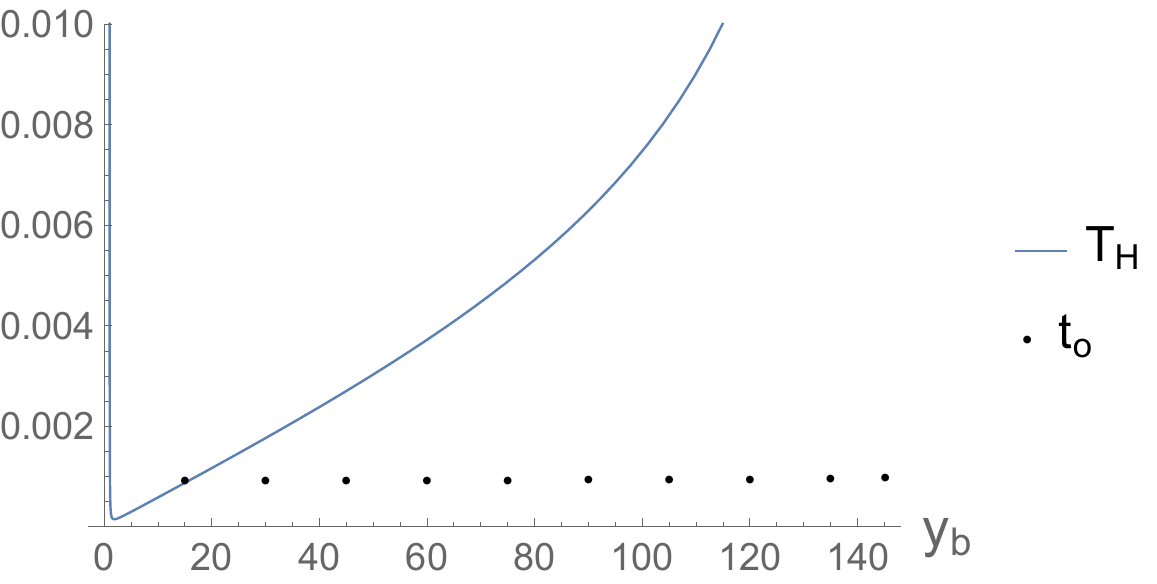}
         &   \includegraphics[height=4cm]{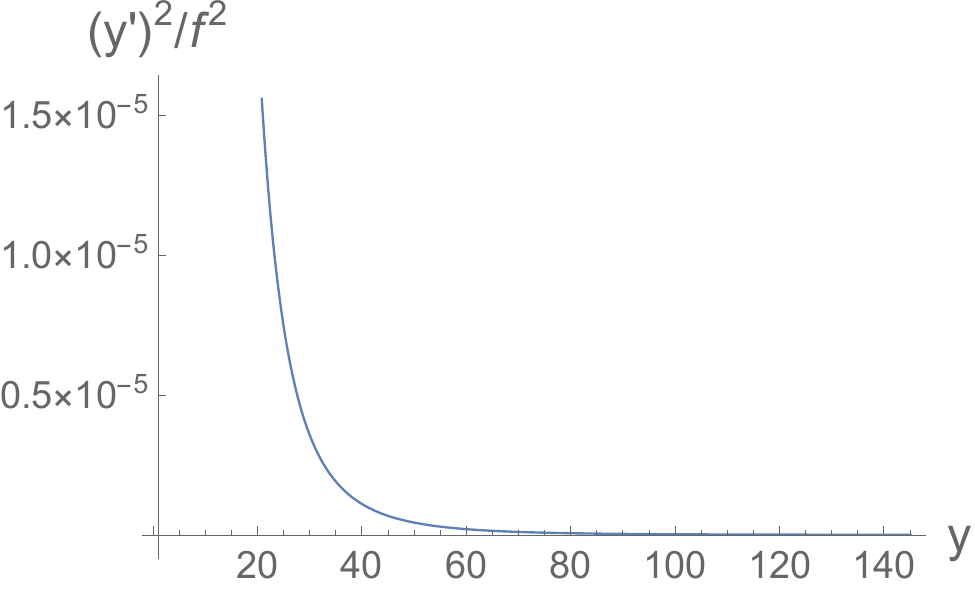}\\
         (a) & (b) &(c)
   \end{tabular}
}
\caption{\textbf{Time scales comparison for data in Table \ref{timescales}.} $d=4$, $L_{AdS}=1$, $r_+=100$ ($\gamma=0.01$), $r_-=99.9$, $l=10$, $T=0.999999999$. (a) Hubble time $T_H$, oscillation time $t_o$ and decay time $t_d$ as a function of the brane radius $y_b$. (b) Detail of Hubble time $T_H$ and oscillation time $t_o$ only. (c) Ratio $[y'(t)]^2/f^2(y)$. The adiabatic approximation is reliable for at least part of the brane trajectory considered, as discussed in Section \ref{gravloc}. The condition $[y'(t)]^2/f^2(y)\ll 1$ is also satisfied.}
\label{timesc1}
\end{figure}

\begin{figure}[H]
\centering
\resizebox{\textwidth}{!}{
    \begin{tabular}{cc}
\includegraphics[height=4cm]{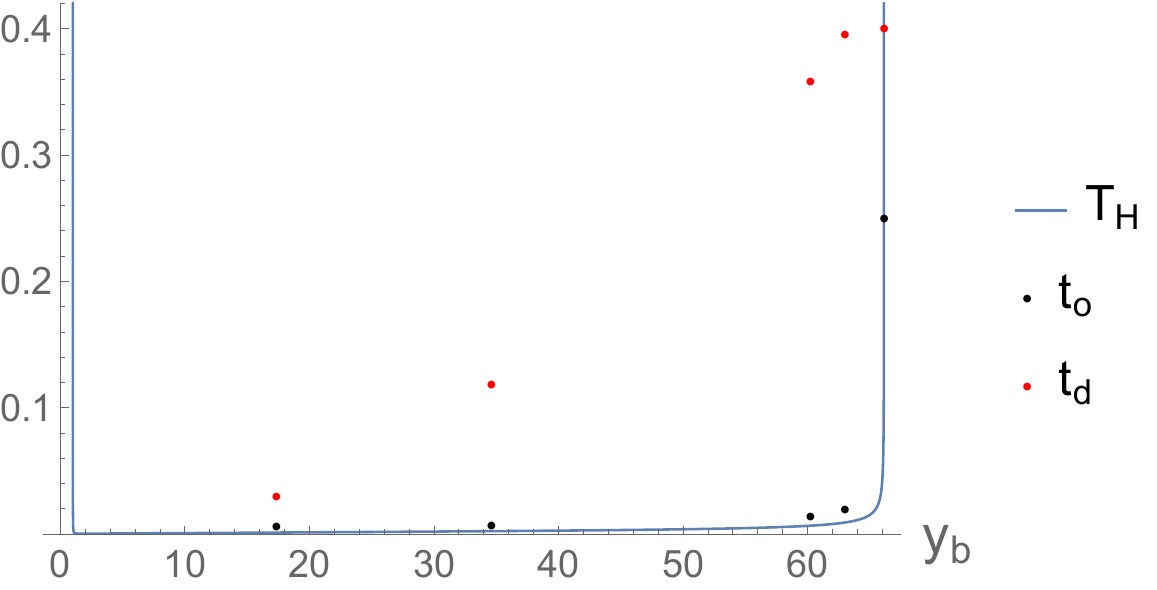}

    &   \includegraphics[height=4cm]{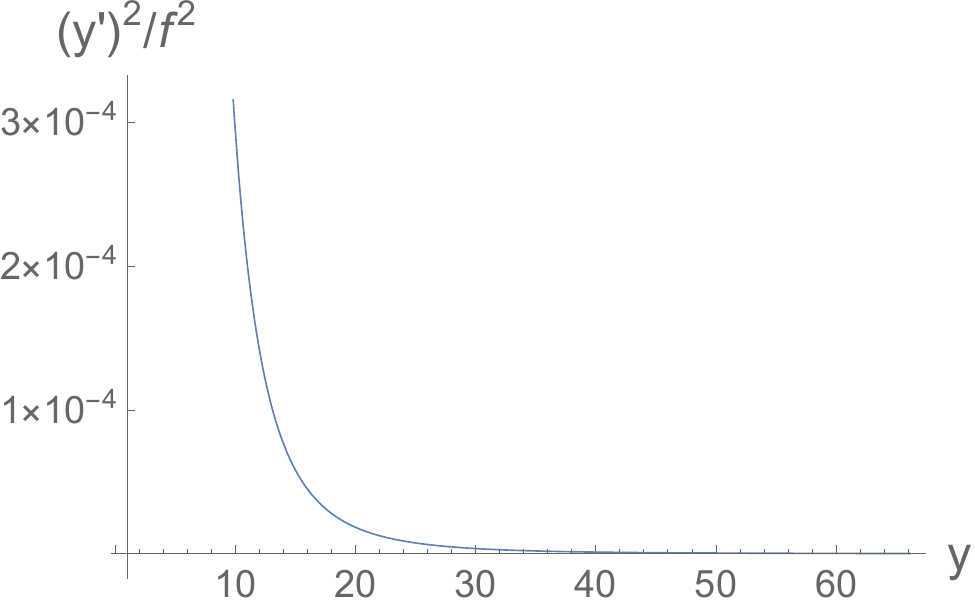}\\
         (a) & (b)
   \end{tabular}
}
\caption{\textbf{Time scales comparison for data in Table \ref{fitresults}.} $d=4$, $L_{AdS}=1$, $r_+=100$ ($\gamma=0.01$), $r_-=99.9$, $l=1$, $T=0.99999993325$. (a) Hubble time $T_H$, oscillation time $t_o$ and decay time $t_d$. (b) Ratio $[y'(t)]^2/f^2(y)$. The adiabatic approximation is clearly not reliable for the parameters considered. The condition $[y'(t)]^2/f^2(y)\ll 1$ is still satisfied.}
\label{timesc2}
\end{figure}

As we pointed out in Section \ref{gravloc}, the three time scales that we must compare in order to verify that the adiabatic approximation we used is reliable are the oscillation time $t_o=1/\bar{\omega}$, the decay time $t_d=2/\Gamma$ and the quantity
\begin{equation}
T_H=\frac{y(t)}{y'(t)}=\frac{\tilde{T}y^2}{f(y)\sqrt{\tilde{T}^2y^2-f(y)}},
\end{equation}
which in comoving coordinates would correspond to the Hubble time of the braneworld cosmology, and that we will call for semplicity ``Hubble time''. If $t_o\ll T_H$ for a given radius of the brane $y=y_b$, the adiabatic approximation holds. We remark that the Hubble time does not depend on the angular momentum $l$. Thus, since increasing $l$ the real part of the frequency increases and the oscillation time $t_o$ decreases, for higher values of $l$ it is easier to satisfy the adiabatic approximation condition. Additionally, as we have explained, the relation
\begin{equation}
\frac{[y'(t)]^2}{f^2(y)}=\frac{\tilde{T}^2y^2}{\tilde{T}^2y^2-f(y)}\ll 1
\label{approx}
\end{equation}
must hold as well at $y=y_b$. We remind that $\tilde{T}=r_+T$. In Section \ref{gravloc} we have shown that, for $l=10$, $\gamma=0.01$, $r_+=100$, $r_-=99.9$ and $T=0.999999999$ it is possible to obtain $t_d\gg T_H\gg t_o$ for a large part of the brane trajectory, meaning that the adiabatic approximation is reliable and that it breaks down (and therefore our analysis loses its meaning) before the graviton leaks into the bulk. Additionally, the condition (\ref{approx}) is always satisfied when the adiabatic approximation holds. For completeness we report in Figure \ref{timesc1} a plot of these quantities as a function of the position of the brane for the data reported in Table \ref{timescales}.
\noindent For the same size and charge of the black hole but with $l=1$ and $T=0.99999993325$ (the maximum radius of the brane is $y_0=r_0/r_+=66.15$), using the results reported in Table \ref{fitresults}, we find that the adiabatic approximation is never reliable (i.e. $T_H<t_o$) during the trajectory of the brane (see Figure \ref{timesc2}). It clearly holds at the turning point $y_0$, where the Hubble time diverges, but there the quasi-bound mode is extremely short-lived ($\bar{\omega}$ and $\Gamma$ are of the same order). The condition (\ref{approx}) is still satisfied for a large part of the brane trajectory. This problem sums up with the discordance between the expected 4-dimensional value of the frequency $\omega_{GR}$ and the one obtained numerically in the $l=1$ case. Even if for smaller black holes (for instance in the $\gamma=0.1$ case) close to the turning point the adiabatic approximation can hold also for $l=1$, we can conclude that an effective 4-dimensional description of gravity localized on the brane is obtained more easily and for a larger amount of brane proper time when the value of the angular momentum $l$ is higher. This result, together with the observation that our analysis is valid only on a time scale smaller than the Hubble time and that the quasi-bound mode is not stable, confirms our expectation to find gravity localization only locally and for a finite amount of time.

\end{appendices}

\addcontentsline{toc}{section}{References}

\printbibliography
\end{document}